\documentstyle[aps,eqsecnum,floats]{revtex}
\begin{document}
\draft
\twocolumn[\hsize\textwidth\columnwidth\hsize\csname 
           @twocolumnfalse\endcsname
\title{Measuring black-hole parameters and testing general
       relativity using gravitational-wave data from
       space-based interferometers}
\author{Eric Poisson$^*$}
\address{Department of Physics, University of Guelph, Guelph,
         Ontario, N1G 2W1, Canada$^\dagger$; \\
         McDonnell Center for the Space Sciences, Department of
         Physics, Washington University, St.~Louis, Missouri,
         63130}
\date{Submitted to Physical Review D, June 12, 1996}
\maketitle

\begin{abstract}
\widetext
Among the expected sources of gravitational waves for the Laser 
Interferometer Space Antenna (LISA) is the capture of solar-mass 
compact stars by massive black holes residing in galactic centers. 
We construct a simple model for such a capture, in which the compact 
star moves freely on a circular orbit in the equatorial plane of the 
massive black hole. We consider the gravitational waves emitted during 
the late stages of orbital evolution, shortly before the orbiting mass 
reaches the innermost stable circular orbit. We construct a simple model 
for the gravitational-wave signal, in which the phasing of the waves 
plays the dominant role. The signal's behavior depends on a number of 
parameters, including $\mu$, the mass of the orbiting star, $M$, the 
mass of the central black hole, and $J$, the black hole's angular momentum.
We calculate, using our simplified model, and in the limit of large 
signal-to-noise ratio, the accuracy with which these quantities can be 
estimated during a gravitational-wave measurement. For concreteness we 
consider a typical system consisting of a $10\ M_\odot$ black hole orbiting 
a nonrotating black hole of mass $10^6\ M_\odot$, whose gravitational 
waves are monitored during an entire year before the orbiting mass reaches 
the innermost stable circular orbit. Defining $\chi \equiv cJ/GM^2$ and 
$\eta \equiv \mu/M$, we find $\Delta \chi \simeq 5 \times 10^{-2}/\rho$, 
$\Delta \eta/ \eta \simeq 6 \times 10^{-2}/\rho$, and 
$\Delta M/M \simeq 2 \times 10^{-3}/\rho$. Here, $\rho$ denotes the 
signal-to-noise ratio associated with the signal and its measurement. 
That these uncertainties are all much smaller than $1/\rho$, the 
signal-to-noise ratio level, is due to the large number of wave cycles 
received by the detector in the course of one year. These are the main 
results of this paper. Our simplified model also suggests a method for 
experimentally testing the strong-field predictions of general relativity.
\end{abstract}
\vskip 2pc]

\narrowtext

\section{Introduction and summary}

\subsection{Space-based gravitational-wave detectors}

The Laser Interferometer Space Antenna (LISA) 
project \cite{LISA} 
was proposed in December 1993 as a cornerstone project
for the ``Horizon 2000 Plus'' program of the European
Space Agency. If the project is realized, LISA might
start searching for gravitational waves in the year
2010.

The LISA interferometer consists of three pairs of spacecrafts, 
each pair located at the corners of an equilateral triangle. 
The constellation moves on a heliocentric orbit at 1 AU, following the 
Earth by an angle of $20^\circ$. Each side of the triangle has a 
length of $5 \times 10^6\ \mbox{km}$, and the plane of the triangle 
is oriented at $60^\circ$ with respect to the ecliptic. 

LISA uses laser interferometry 
to detect gravitational waves in the
frequency band between 0.1 and 100 mHz. Such waves
have much too low a frequency to be measurable by
Earth-based interferometers, which are severely limited 
by seismic noise at these frequencies \cite{LIGO}. 
LISA's projected sensitivity is such that 
a gravitational wave of amplitude $10^{-23}$ would
give rise to a signal-to-noise ratio of 5 after one
year of observation. 

Among the expected sources of gravitational waves
relevant to LISA are 
solar oscillations \cite{CutlerLindblom}, 
binary systems of normal and compact stars 
within the galaxy \cite{Hilsetal}, massive black-hole 
binaries in other galaxies \cite{LISA}, and primordial 
gravitational waves \cite{CaldwellAllen}. 

In this paper we consider another type of source, also 
expected to be highly relevant: the capture of solar-mass 
compact stars by massive black holes 
\cite{HilsBender,Shibata}.

\subsection{Stellar capture by massive black holes}

The most likely explanation for active galactic nuclei
involves the tidal disruption of normal stars by a massive
black hole residing in the galactic center. It is
plausible that compact stars (such as neutron
stars and black holes) will also be captured by the
massive black hole. However, such stars will not be disrupted,
and their complex orbital motion will generate gravitational
waves \cite{HilsBender}. 
If the central black hole has a mass in the interval
between $10^5$ and $10^7$ solar masses, then these waves
will have their frequency within the LISA 
bandwidth \cite{LISA}, and
will be detected if they are sufficiently strong.

It is expected that typically, a captured star will 
move on a highly eccentric orbit around the massive black 
hole \cite{HilsBender,Shibata,Tanakaetal,CKP}. Such orbits
are easily perturbed by the other stars in the galactic nucleus,
and this can have a damaging effect on the gravitational-wave
signal. The probability for an orbit to be perturbed decreases
with increasing values of the orbiting mass. For this reason 
it is judged \cite{LISA,HilsBender2}
that the capture of compact stars with masses in the 
interval between 5 and 10 solar masses would be the most 
relevant for observations by LISA. Compact stars of such
masses cannot be neutron stars, and would therefore be
black holes. It was estimated \cite{LISA,HilsBender2}
that the rate of occurrence for such events could be as 
large as several per year, and that the signal-to-noise 
ratio could be near 40 for a capture occurring at a distance 
corresponding to a cosmological redshift of one. 

In this paper we consider the gravitational waves produced
during the capture of a $10\ M_\odot$ black hole by a
massive, $10^6\ M_\odot$ black hole. We are specifically
interested in what can be learned from observing
these waves.

\subsection{Information carried by the waves}

The gravitational waves produced during the capture of a
compact star by a massive black hole are rich in information.
In principle, measuring the waves should allow us to extract 
the value of the black-hole parameters (mass and angular 
momentum), as well as to test the strong-field predictions of 
general relativity \cite{RyanMM}. In this paper we wish to 
propose a strategy for doing both. 

A possible method for extracting the black-hole parameters,
which we shall not explore in this paper, would go as 
follows \cite{FinnOriThorne}.

The gravitational waves generated by the bound motion of a 
mass around a center come with a certain distribution
in frequency. For example, circular motion in the equatorial
plane of a black hole produces waves with
frequencies at every harmonic of the orbital 
frequency \cite{PoissonI}.
(The strongest waves come at twice the orbital frequency.)
As an other example, the gravitational waves produced by 
a mass in eccentric motion (also in the equatorial
plane) come in harmonics of two fundamental 
frequencies \cite{Shibata,CKP,Tanakaetal}, 
respectively associated with the angular and radial components 
of the motion. 

Precisely how the waves are distributed in frequency depends 
on the orbital parameters (for example, the eccentricity). 
Measuring the distribution at some moment in time
therefore leads to an estimation of these parameters.
Now, because the system is losing energy and angular 
momentum to gravitational waves,
the orbital parameters are not constant, but evolve
in time. Precisely how they do so depends on the
black-hole parameters (and also on the mass of the
orbiting object). Measuring how the orbital
parameters evolve in time therefore leads to an 
estimation of the black-hole parameters.

Such a method was recently examined by Finn, Ori, and
Thorne \cite{FinnOriThorne}, in the restricted context
of circular, equatorial orbits.  
The method involves measuring the relative 
amplitude of each frequency component of the 
gravitational waves. In Sec.~V we will see
that amplitude measurements typically come with
relative statistical uncertainties of order
$1/\rho$, where $\rho$ is the signal-to-noise
ratio associated with the signal and its
measurement. (The uncertainty is due to the
detector noise.) It follows that by using this
method, the estimation of the black-hole parameters 
can be effected with a relative accuracy of order $1/\rho$. 

Another method for extracting the black-hole parameters, 
also not explored in this paper, would involve monitoring 
that part of the gravitational-wave signal which corresponds
to the black hole settling down after the orbiting mass'
final plunge \cite{Echeverria,Finn}. 
This part of the signal is strongly
dominated by the black-hole 
quasi-normal modes \cite{QNM}, whose
frequencies and damping times are functions of 
the hole's  mass and angular momentum. Such a method 
was examined by Echeverria \cite{Echeverria} and 
Finn \cite{Finn}, who also conclude 
that the black-hole parameters can be estimated 
with a relative accuracy of order $1/\rho$.

The method considered in this paper is essentially
due to Finn and Chernoff \cite{FinnChernoff}, and
to Cutler and Flanagan \cite{CutlerFlanagan}. It
goes as follows.

As was pointed out above, the gravitational waves come 
in many frequency components. One of these, preferably the 
dominant one, is selected, and its phase is carefully
monitored. Let us denote by $f$ the frequency of this
component. Because the system loses energy and angular
momentum to gravitational waves, $f$ increases slowly 
with time. The phasing of the waves is determined by $df/dt$,
the rate of change of the frequency. In turn, this depends on 
the black-hole parameters (and also on the mass of the orbiting 
object). Monitoring the phase therefore leads to a way of 
estimating the black-hole parameters.

We will show in Sec.~V that the estimation of those parameters 
which influence the phasing of the waves  
can be effected with a relative accuracy 
of order $1/\rho_{\rm eff}$, where the {\it effective}
signal-to-noise ratio is given by $\rho_{\rm eff}
= 2\pi {\cal N} \rho$. Here, $\cal N$ is the
total number of wave cycles received by the
gravitational-wave detector. This number will
be estimated in Sec.~III: For a binary system 
consisting of two black holes, one of mass
$10\ M_\odot$, the other of mass $10^6\ M_\odot$, 
and for a signal monitored during an entire year,  
${\cal N} \simeq 10^5$. The method considered in this paper 
therefore leads to a much enhanced accuracy 
for the estimation of the black-hole parameters. 

\subsection{Realistic captures and our simplified model}

The realistic description of a capture is complicated. The
captured star moves on a highly eccentric 
orbit \cite{HilsBender,Shibata,Tanakaetal,CKP} which 
may not lie in the equatorial plane of the massive black hole. 
The orbital parameters (such as periastron and eccentricity) 
evolve in time, in part stochastically (due to 
perturbation by other stars) and in part secularly 
(due to gravitational radiation reaction).

To model a realistic capture, and to calculate the gravitational 
waves emitted, is a difficult problem. To a first approximation, 
in which one ignores the external perturbation and the radiation 
reaction, the orbit is accurately 
represented by a bound geodesic of the Kerr 
spacetime \cite{MTW,BPT}. Such an 
orbit is characterized by three parameters: orbital energy, 
orbital angular momentum, and a third called the Carter constant.

An improved approximation would incorporate a small stochastic 
variation of these parameters, which would account for the 
external perturbation, and also a small secular variation, 
which would account for the radiation 
reaction \cite{Shibata,Tanakaetal,CKP}. This approximation
should be quite accurate if the density of stars is sufficiently 
low near the central black hole. 

The main difficulty preventing this problem from finding a solution 
resides in the fact that no prescription is currently available for 
calculating the secular evolution of the Carter 
constant. (See, however, Refs.~\cite{Shibataetal,Ryan,Ori}.
The evolution of the other orbital parameters is
obtained by computing how much energy and angular momentum 
are carried off by the gravitational waves. This is done 
by applying the methods of black-hole perturbation 
theory \cite{Shibata,Tanakaetal,CKP}.) 
However, because of the system's small mass ratio,
this problem should be much more tractable than the 
nonrestricted general relativistic two-body problem, 
which currently is only amenable to post-Newtonian 
methods \cite{Will}.

In this paper we choose to simplify the problem as much
as possible, in order to introduce a simple, analytic expression
for the gravitational-wave signal. We shall assume that the 
captured object moves freely on a circular orbit in the equatorial 
plane of the massive black hole. We shall ignore all external 
perturbations, and focus on the secular evolution induced by 
the emission of gravitational waves. Because the orbit is 
circular and equatorial, this evolution is dictated entirely 
by the loss of orbital energy. And because the ratio of the 
two masses is small, this loss is accurately 
computed using the methods of black-hole perturbation 
theory \cite{Shibata2}.

\subsection{Model gravitational-wave signal}

As the system loses energy and angular momentum to gravitational 
waves, $f$, the frequency of the selected gravitational-wave
component, slowly increases. For a circular orbit, the dominant
component is at twice the orbital frequency; if $\Omega$
denotes the angular velocity of the orbiting mass, then
$2\pi f = 2 \Omega$. It can be shown \cite{CKP} that 
during most of the orbital evolution, the {\it adiabatic 
approximation} holds. This states that the radiation reaction 
timescale is much longer than the orbital period, so that
$df/dt \ll 1/f^2$. Our analysis relies heavily on the
adiabatic approximation. 

There exists an upper bound for the gravitational-wave frequency,
which we denote by $f_0$. This bound is set by the innermost stable 
circular orbit (ISCO) of the Kerr spacetime, and
is equal to $\Omega_0/\pi$, where $\Omega_0$ is the angular
velocity of a test mass moving on the ISCO. The gravitational 
waves therefore sweep up in frequency until $f=f_0$. At this 
point, the orbiting mass can no longer maintain a circular orbit,
and is forced to plunge into the massive black hole, thereby
shutting off the gravitational waves. It should be noted that 
the adiabatic approximation necessarily breaks down as the 
orbiting mass reaches the ISCO \cite{CKP}. 

We shall focus our attention 
on that part of the orbital evolution for which the
gravitational waves are strongest and most interesting.
This corresponds to the very late stages, shortly before the 
orbiting mass reaches the ISCO. If we assume that
$f$ is always very close to $f_0$, then we may express 
all relevant quantities as Taylor expansions about 
the ISCO. 

We are mostly interested in the phasing of the waves,
which is determined by $df/dt$.  
We will find in Sec.~II B that this can be expressed as
\begin{equation}
\frac{df}{dt} = \frac{\pi}{3}\, 
\frac{{\cal V} {f_0}^2}{1-f/f_0}\, 
\Bigl[ 1 - 2\alpha\,(1-f/f_0) + \cdots \Bigr].
\label{1.1}
\end{equation}
Here, ${\cal V}$ and $\alpha$ are dimensionless expansion 
coefficients which depend on $\mu$, the mass 
of the orbiting star, $M$, the mass of the central black hole, 
and $|\chi| \equiv J/M^2$, its dimensionless angular-momentum 
parameter. (We use units such that $G=c=1$; $J$ denotes the 
black hole's spin angular momentum, and $\chi$ will be defined 
more precisely in Sec.~IV.) The dots denote terms of higher
order in $1-f/f_0$. It is to be noted that Eq.~(\ref{1.1}) 
is singular at $f=f_0$. This reflects the breakdown of 
the adiabatic approximation at the ISCO. 
This issue is discussed in detail in Sec.~II C.

We use Eq.~(\ref{1.1}) to construct a model for the 
gravitational-wave signal originating from
the capture of a compact star by a massive black hole.
The model signal then depends on the parameters $f_0$,
${\cal V}$, and $\alpha$, which codify information about 
the source. We ask how well these parameters
can be estimated during a gravitational-wave measurement,
and what they can tell us about $\mu$, $M$, and $\chi$.

In the following we will refer to $\{f_0, {\cal V}, 
\alpha\}$ as the {\it signal} parameters, and to
$\{\mu,M,\chi\}$ as the {\it source} parameters. 

\subsection{A remark on our expansion approach}

The expansion (\ref{1.1}) is accurate only if 
$1-f/f_0$ is always much smaller than unity. 
We must ask whether this is true for a
system with $\mu = 10\ M_\odot$, $M= 10^6\ M_\odot$,
whose gravitational waves are monitored for an entire
year before the orbiting mass reaches the innermost
stable circular orbit. The answer, most unfortunately,
is no. For such a system, as we shall see in Sec.~III, 
$\varepsilon \equiv \Delta f/f_0 \simeq 1.1$, where
$\Delta f$ is the frequency interval over which the 
signal is monitored.

To build a model signal upon such expansions as 
Eq.~(\ref{1.1}) therefore seems a bad idea if one
is at all interested in accuracy. Should we then give up
this approach and do something better? We choose to
give a negative answer: We will not give up, and we 
shall proceed with our expansions, even though the 
resulting signal may not be very accurate. 

The reason is that to provide a better representation
for $df/dt$ would require a great deal more work, and 
could not be done in the simple, analytic way advocated
in this paper. Indeed, the gravitational-wave signal 
would have to be generated numerically. Now, numerical 
signals {\it will} have to be provided when an attempt 
is made to construct more sophisticated models, for example, 
models that incorporate highly eccentric orbits. 
However, we feel it would not be worthwhile to introduce
numerical signals at this stage.

Instead, for the sake of simplicity, we shall continue to
treat $1-f/f_0$ as a formally small quantity, and to carry out 
expansions in powers of that quantity. We hope that our 
conclusions will not depend strongly on this assumption, 
and that we can safely put $\varepsilon \simeq 1$ at the 
end of the calculation.

\subsection{The main results}

Our model gravitational-wave signal is characterized
by the parameters $f_0$, ${\cal V}$, and $\alpha$,
which are functions of the source parameters 
$\mu$, $M$, and $\chi$.

The signal parameters can be estimated during a gravitational-wave 
measurement. The statistical uncertainties associated with such
an estimation are calculated in detail in Sec.~V,
in the limit of large signal-to-noise ratio. For concreteness
we consider a typical system consisting of a $10\ M_\odot$ black 
hole orbiting a nonrotating black hole of mass $10^6\ M_\odot$, 
whose gravitational waves are monitored for an entire year before 
the orbiting mass reaches the innermost stable circular orbit. 
For such a system we find that the measurement uncertainties 
are given by
\begin{eqnarray}
\frac{\Delta f_0}{f_0} &\simeq& 5.5 \times 10^{-4}/\rho,
\nonumber \\
\frac{\Delta {\cal V}}{{\cal V}} &\simeq&
3.8 \times 10^{-3}/\rho,
\label{1.2} \\
\Delta \alpha &\simeq& 1.0 \times 10^{-2}/\rho,
\nonumber
\end{eqnarray}
where $\rho$ is the signal-to-noise ratio. A more precise,
and more general, statement of these results appears in 
Eq.~(\ref{5.11}). These uncertainties should be compared 
with that associated with $\cal A$, 
the signal's amplitude parameter: 
$\Delta {\cal A}/ {\cal A} = 1/\rho$. That the phase 
parameters can be estimated much more accurately is due 
to the large number of wave cycles received by the 
detector. We indeed recall that ${\cal N} \sim 10^5$.

Equations (\ref{1.2}) imply that the source parameters
$\chi$, $\eta \equiv \mu/M$, and $M$ can be determined 
with uncertainties
\begin{eqnarray}
\Delta \chi &\simeq& 5.0 \times 10^{-2}/\rho,
\nonumber \\
\frac{\Delta \eta}{\eta} &\simeq& 5.5 \times 10^{-2}/\rho,
\label{1.3} \\
\frac{\Delta M}{M} &\simeq& 2.4 \times 10^{-3}/\rho.
\nonumber
\end{eqnarray}
A more precise, and more general, statement of there results
can be found in Sec.~V C. We recall from Sec.~I B that 
gravitational waves produced during a capture occurring at 
cosmological distances could be measured with $\rho \simeq 40$.  

\subsection{Testing general relativity}

In order to obtain Eqs.~(\ref{1.3}) from (\ref{1.2}),
the relation between the source parameters $\{\mu,M,\chi\}$
and the signal parameters $\{f_0,{\cal V},\alpha\}$ must
be known. This relation is based upon two conditions: that 
the central mass is a Kerr black hole, and that the system
emits gravitational waves according to the quantitative 
predictions of Einstein's theory. Indeed, general relativity 
is used to calculate $dE/dt$, the energy lost per unit time 
to gravitational waves. It is also used to write $f=\Omega/\pi$,
the statement that the strongest waves come at twice the orbital
frequency. On the other hand, the Kerr metric is used to calculate 
$dE/d\Omega$, the relation between orbital energy and angular 
velocity. Putting all this together gives $df/dt$ in terms of the
source parameters.

It is easy to imagine a situation in which either one,
or both, of these conditions would be violated. For
example, the central mass could be something quite
different from a Kerr black hole, or Einstein's theory
could make wrong predictions regarding the emission
of gravitational waves. As a specific example, we may
recall that in the Brans-Dicke theory \cite{BransDicke}, 
the existence of scalar radiation implies a different 
result for $dE/dt$ from what is predicted by general 
relativity \cite{WillBD}. However, by 
virtue of the uniqueness of the Kerr black 
hole \cite{Hawking}, the relation $dE/d\Omega$ 
is preserved in the Brans-Dicke theory.

A breakdown of what we shall call the standard
model (general relativity is valid and the central mass
is a Kerr black hole) does not logically invalidate
Eq.~(\ref{1.1}). Indeed, this equation is based 
solely on the assumptions that the orbiting
mass moves on a circular orbit, and that the spacetime
possesses an innermost stable circular orbit. These conditions 
could easily be preserved in an alternative model. 

The standard model implies a particular relationship
between the source and signal parameters. Another
model would presumably imply a different 
relationship. We may therefore ask whether a measurement
of the signal parameters might allow us to test the
validity of the standard model. This question is
considered in Sec.~V D, where we show that a measurement
of $\alpha$ outside the interval $(-\infty,1.8)$ would
unambiguously signal the breakdown of the standard model.

\subsection{Conclusion}

The analysis presented in this paper is by no means
definitive, and needs to be improved in many respects. 
First, the crude representation (\ref{1.1}) 
for $df/dt$ will have to be replaced by something more 
reliable, presumably of numerical form. Second, 
the unrealistic assumption that the orbit must be 
equatorial and circular will have to be removed. All
of this will require a great deal of effort.

However, our simplified analysis should not be criticized 
too harshly. It is, after all, a first approach to a difficult 
problem. And our conclusions show promise: We believe that 
our results (\ref{1.3}) for the measurement accuracies ---  
most especially the fact that the phasing method yields 
uncertainties which are much smaller than $1/\rho$ --- provide 
ample motivation for pursuing this work and adding 
new layers of complexity. 

\subsection{Organization of this paper}

We begin in Sec.~II with a detailed introduction of our 
model for the gravitational-wave signal. The assumptions 
are reviewed and explained in Sec.~II A. The signal is
explicitly constructed in Sec.~II B. Finally, some of the 
approximations employed are motivated in Sec.~II C. 

In Sec.~III we consider a specific case of the general
framework introduced in Sec.~II, in which the central
mass is a Schwarzschild black hole. This
section also provides numerical estimates for such
quantities as $f_0$, ${\cal V}$, $\alpha$, 
$\varepsilon$, and ${\cal N}$. 

In Sec.~IV we generalize the discussion of Sec.~III to
the case of a Kerr black hole.

Section V contains a discussion of the statistical
uncertainties associated with the estimation of the
signal parameters. In Sec.~V A we re-introduce the 
model signal. In Sec.~V B, expressions are derived for the 
measurement uncertainties.  These are converted, in Sec.~V C, 
into uncertainties for the source parameters,
assuming the validity of the standard model. Finally,
in Sec.~V D, a method is proposed to test the validity
of the standard model.

The Appendix contains a general discussion of orbital
motion in spacetimes which are stationary, asymptotically flat, 
axially symmetric, and reflection symmetric about the
equatorial plane. (The orbits are restricted to this 
plane.) The equations of motion are introduced 
in Sec.~A 1. Circular orbits are defined precisely in 
Sec.~A 2. In Sec.~A 3 we consider two neighboring circular 
orbits, and derive relationships between the displacements 
in radius, energy, angular momentum, and angular velocity. 
Finally, the innermost stable circular orbit is defined 
precisely in Sec.~A 4.

\section{Model gravitational-wave signal}

In this section we introduce our model for the gravitational
waves emitted during the late stages of orbital evolution of a 
binary system with small mass ratio, shortly before the smaller
mass reaches the innermost stable circular orbit associated
with the larger mass. In Sec.~II A we list and explain the 
assumptions involved in formulating the model. In Sec.~II B 
we construct the signal. In Sec.~II C we discuss the 
validity of some of the approximations used in subsection B.

\subsection{Assumptions}

To formulate the model we shall make a number of assumptions.
These are listed and explained below.

{\it Assumption 1:} The binary system consists of a small mass 
$\mu$ and a large mass $M$, so that $\mu/M \ll 1$. 
In typical situations we shall take $\mu = 10\, M_\odot$ 
and $M = 10^6\, M_\odot$. 

{\it Assumption 2:} The spacetime geometry associated with the 
mass $M$ is stationary, asymptotically flat, axially 
symmetric, and reflection symmetric about the equatorial plane. 
However, it is not assumed that $M$ is a Kerr black hole, or 
even that the metric is a solution to the Einstein field equations. 
Although most of this paper will be devoted to 
the specific case of a Kerr black hole, we wish to keep the 
framework as general as possible in order to later identify a 
strategy for experimentally testing  the predictions of 
general relativity.

Assumption 2 leads to a number of simplifications.
First, the fact that the spacetime is stationary and axially 
symmetric implies that $\Omega=d\phi/dt$, the angular velocity 
of the orbiting mass $\mu$, is uniform if the orbit is circular. 
Here, $\phi$ is the angle that goes around the axis of symmetry, 
and $t$ is proper time for 
observers at rest at infinity. Second, the fact that the 
spacetime is reflection symmetric about the equatorial plane 
implies that if started in this plane, the orbit will remain in
the plane forever (whether it is circular or not). Consequently,
we can further impose:

{\it Assumption 3:} The mass $\mu$ moves on a 
circular orbit in the equatorial plane of the 
mass $M$. This assumption is imposed solely for the 
sake of simplicity, and cannot be expected to hold in 
realistic situations. This was discussed in detail
in Sec.~I D.

{\it Assumption 4:} The spacetime possesses an innermost
stable circular orbit (ISCO). The ISCO is defined to be
that particular circular orbit for which an infinitesimal
change in orbital energy induces a large change in the
orbital radius, or in the angular velocity $\Omega$. 

The ISCO can be thought of as the limiting member of 
a sequence of circular orbits, each labeled by the 
value of its angular velocity. For $\Omega$ smaller 
than the critical value $\Omega_0$, stable circular 
orbits exist; for $\Omega$ larger than $\Omega_0$, 
circular orbits are no longer mechanically stable. 
Bound orbits within the ISCO all plunge toward
the mass $M$. 

{\it Assumption 5:} The binary system emits gravitational
waves. It is not assumed that Einstein's theory gives the
correct description of gravitational radiation, for the 
same reason as was expressed in the context of assumption
2. 

Due to the emission of gravitational waves, the system 
loses energy and angular momentum, and as a result, 
the mass $\mu$ undergoes orbital evolution: its angular 
velocity does not stay constant but keeps on increasing 
as the orbital radius decreases. 

Combining assumptions 1 and 5 leads to the consequence
that while $\Omega$ increases due to the radiation
reaction, this increase occurs 
{\it adiabatically} \cite{CKP}: the timescale for the
increase is much larger than the orbital period.
The orbital evolution can therefore
be represented by a succession of circular orbits 
(ordered by the increasing value of the angular velocity),
with the mass $\mu$ moving quasi-statically from
one orbit to the next. 

It is important to state that the adiabatic approximation 
must break down as the orbiting mass reaches the innermost 
stable circular orbit \cite{CKP}. This follows from the very 
definition of the ISCO. Nevertheless, the adiabatic approximation 
holds for most of the orbital evolution, and its 
breakdown will not be an obstacle in the forthcoming analysis. 

{\it Assumption 6:} The gravitational waves emitted by
the binary system are monitored by a space-based interferometric
detector (such as LISA) during an entire year before the mass
$\mu$ reaches the innermost stable circular orbit.

We shall calculate that during this time, the angular 
velocity changes by a fractional amount of order unity. We 
shall also find that $\cal N$, the total number of wave 
cycles received by the detector during the year's worth 
of observation, is quite large. 
In typical situations, ${\cal N} \sim 10^5$. 

\subsection{Construction of the signal}

In brief, the model gravitational-wave signal is based upon a Taylor 
expansion about the innermost stable circular orbit. The justification
for this was given in Sec.~I F, and will be repeated in Sec.~II C.
The model signal is also based upon the adiabatic approximation, 
which is itself a consequence of assumptions 1 and 5. 
And finally, we shall construct the signal by focusing mostly
on its {\it phasing}, as was motivated in Sec.~I C.

We now proceed. Combining the adiabatic approximation with 
assumption 3 (regarding the circularity of the orbit) 
implies that the gravitational waves have a frequency 
component at {\it every} harmonic of the angular velocity 
$\Omega(t)$. (Here we indicate that the angular velocity 
evolves slowly with time.) More precisely \cite{CKP,PoissonI}, 
the frequency $F_m$ associated with 
the $m$'th harmonic is given by  $2\pi F_m(t) = m\,\Omega(t)$. 
We single out the strongest frequency component, 
and assume that only this component is actually 
measured by the gravitational-wave detector. 
(In general relativity the dominant harmonic 
is at $m=2$.) This particular frequency will be 
denoted by $F(t)$.

The phasing of the wave is determined by the function 
$F(t)$. To obtain this we invoke once again 
the adiabatic approximation, and write
\begin{equation}
\frac{dF}{dt} = \frac{dE/dt}{dE/dF},
\label{2.1}
\end{equation}
where $E$ denotes the orbital energy of
the mass $\mu$. The numerator of Eq.~(\ref{2.1})
represents the energy lost per unit time to gravitational waves. 
The denominator gives the relationship between
orbital energy and angular velocity. The adiabatic approximation
ensures that $E$ and $F$ change very little over 
the course of one orbit, so that these quantities are actually
well defined. As was pointed out before, the adiabatic 
approximation, and therefore Eq.~{(2.1)} also, break down
as the mass $\mu$ reaches the innermost stable circular orbit. 
This does not represent an obstacle to our modeling; we will 
return to this point in Sec.~II C. 

We denote by $f_0$ the gravitational-wave frequency at the
innermost stable circular orbit. To indicate that $f_0$
must scale as the inverse power of $M$,
we introduce the notation
\begin{equation}
\pi M f_0 \equiv {v_0}^3,
\label{2.2}
\end{equation}
where $v_0$ is a scale-invariant quantity; the factor of $\pi$
was inserted for convenience.

We expand $dE/dt$ about the innermost stable circular orbit. 
Since the gravitational-wave luminosity must scale as the 
square of the mass ratio, we write
\begin{equation}
\frac{dE}{dt} = - \Bigl( \frac{\mu}{M} \Bigr)^2\,
{v_0}^{10}\, p\, \bigl[ 1 - \hat{p}\, (1-F/f_0) + \cdots \bigr].
\label{2.3}
\end{equation}
Here, $p$ and $\hat{p}$ are phenomenological parameters.
The value that these parameters take depends on the
nature of the mass $M$ and on the correct description of
gravitational radiation. Specific cases will be 
considered in Sec.~III and VI. For the time being,
however, we leave these parameters undetermined.
The factor of ${v_0}^{10}$ was inserted for convenience, 
and the dots designate terms of higher order
in $1-F/f_0$.

We also expand $dE/dF$ about the innermost stable circular
orbit. To do this we recall that the ISCO is defined to be
that particular orbit for which an infinitesimal change in
orbital energy produces a large change in the orbital
radius, or in the angular velocity (which is related to
the radius by a Kepler-type relation). The defining property 
of the ISCO is therefore that $dE/dF$ must go to
zero there. (See the Appendix for further details, and for
a justification of the fact that $dE/dF$ must vanish 
linearly with $1-F/f_0$.) We also note that $dE/dF$ must 
scale like the product of the two masses, as can be quickly 
seen from an elementary Newtonian calculation. 
We therefore write
\begin{equation}
\frac{dE}{dF} = -\frac{\mu M}{v_0}\, q\, (1-F/f_0)\, 
\bigl[ 1 - \hat{q}\, (1-F/f_0) + \cdots \bigr].
\label{2.4}
\end{equation}
As before, $q$ and $\hat{q}$ are phenomenological parameters,
and a factor of $1/v_0$ was inserted for convenience.

Combining Eqs.~(\ref{2.1})--(\ref{2.4}) yields
\begin{equation}
\frac{dF}{dt} = \frac{\pi}{3}\, {\cal V}\, {f_0}^2\,
(1-F/f_0)^{-1}\, \bigl[ 1 - 2\alpha\,(1-F/f_0) + \cdots \bigr],
\label{2.5}
\end{equation}
where 
\begin{equation}
{\cal V} = \frac{3\pi p}{q}\, \frac{\mu}{M}\, {v_0}^5, 
\qquad
\alpha = {\textstyle \frac{1}{2}} (\hat{p}-\hat{q}).
\label{2.6}
\end{equation}
Equation (\ref{2.5}) can easily be integrated to give
$t(F)$, and the phase function
\begin{equation}
\Phi(F) = \int 2\pi F(t)\, dt = \int 2\pi F\, \frac{dt}{dF}\, dF
\label{2.7}
\end{equation}
can be obtained similarly.

We take the time-domain gravitational-wave signal to be
\begin{equation}
h\bigl(F(t)\bigr) = {\cal A}\, (1+\cdots)\, e^{-i \Phi(F)}.
\label{2.8}
\end{equation}
Here, ${\cal A}$ is the signal's amplitude at the time the 
mass $\mu$ reaches the innermost stable circular orbit; 
${\cal A}$ is inversely proportional to the distance 
to the source. We shall also need an expression for the 
frequency-domain signal, which is obtained by Fourier 
transforming $h(t)$:
\begin{equation}
\tilde{h}(f) = \int h(t)\,e^{2\pi i f t}\, dt =
\int h(F)\, e^{2\pi i f t(F)}\, \frac{dt}{dF}\, dF.
\label{2.9}
\end{equation}
The frequency $f$ must be distinguished from $F(t)$, the 
function of time which was also referred to as the frequency.

We evaluate the Fourier transform by invoking the 
stationary-phase approximation \cite{Jackson}, 
which actually follows from the
adiabatic approximation \cite{CutlerFlanagan}. 
(We shall return to this point
shortly.) After substituting Eq.~(\ref{2.8}) into (\ref{2.9}),
we find that the argument of the exponential becomes
$i$ times $2\pi f t(F) - \Phi(F) \equiv \varphi(F)$. 
Using Eq.~(\ref{2.7}), we see that this phase has a 
stationary point when $0 = d\varphi/dF = 2\pi(f-F)(dt/dF)$, 
that is, when $F=f$. The stationary-phase approximation 
then returns a factor  
$[-\varphi''(f)]^{-1/2} \propto (1-f/f_0)^{-1/2}$ which
must be combined with the factor $dt/dF|_f \propto
(1-f/f_0)$ coming from inside the integral. After simple 
manipulations, including the absorption of constants 
into a redefinition of ${\cal A}$, we arrive at
\begin{eqnarray}
\tilde{h}(f) &=& {\cal A}\, (1 + \cdots)\, (1-f/f_0)^{1/2}\,
e^{i \psi(f)},
\nonumber \\
\psi(f) &=& 2\pi f t_0 - \phi_0 + \phi(f),
\label{2.10} \\
\phi(f) &=& \frac{(1-f/f_0)^3}{{\cal V}}\, \bigl[
1 + \alpha (1-f/f_0) + \cdots \bigr].
\nonumber
\end{eqnarray}
Here, the constants $t_0 \equiv t(f_0)$ and 
$\phi_0 \equiv \Phi(f_0) + \pi/4$ respectively represent
time and phase at the innermost stable circular orbit. As 
before, the dots designate terms of higher order in $1-f/f_0$. 

We shall adopt Eq.~(\ref{2.10}) as our model for the 
(frequency-domain) gravitational-wave signal. It involves six
phenomenological parameters: the amplitude $\cal A$, the
final time $t_0$, the final phase $\phi_0$, the final frequency 
$f_0$, as well as ${\cal V}$ and $\alpha$ coming from 
the expansion of $dF/dt$ about the innermost stable circular orbit.

\subsection{Validity of approximations}

Our model gravitational-wave signal is based
upon a Taylor expansion about the innermost stable circular
orbit. Such an expansion can be accurate only if $1-f/f_0$ 
is small throughout the entire frequency interval over which
the signal is monitored. If $f_i$ denotes the initial value 
of the frequency, we would require $\varepsilon \equiv 
1-f_i/f_0 \ll 1$. 

In Sec.~III we will find that in typical situations, 
$\varepsilon \simeq 1.1$, and is 
therefore not small. This means that our model signal 
cannot be very accurate for frequencies near
$f_i$. This is most unfortunate, but we shall nevertheless proceed 
with our model signal, which at least has the virtue of being 
extremely simple. In view of the fact that our model also 
incorporates unrealistic assumptions 
(that the orbit must be equatorial and 
circular), we feel that this degree of sophistication should be 
sufficient for our purposes. We do not expect that the use,
in our calculations, of a more exact representation for
$dF/dt$ would lead to very different conclusions. 
A more complete discussion of this point was given in Sec.~I F.

We now discuss the breakdown of the 
adiabatic and stationary-phase approximations. The adiabatic 
approximation becomes invalid when the timescale over which
$F(t)$ changes becomes comparable to the orbital period. This
translates to $(1/F)(dF/dt) \sim F$ or, upon using
Eq.~(\ref{2.5}) and $F \sim f_0$,
\begin{equation}
(1-F/f_0) \sim {\cal V}.
\label{2.11}
\end{equation} 
It can be shown that the same criterion applies to
the breakdown of the stationary-phase 
approximation \cite{CutlerFlanagan}. 

To circumvent the problems associated with the breakdown of these 
approximations, we shall impose that Eq.~(\ref{2.10}) is valid for
\begin{equation}
f < f_f \equiv f_0\, (1-{\cal V})
\label{2.12}
\end{equation}
only. For $f>f_f$, the signal is taken to be identically zero.
In any event, the part of the signal 
for which Eq.~(\ref{2.12}) is violated turns out to be 
irrelevant for the purpose of Sec.~V. 
Indeed, we shall see that most of the signal-to-noise
ratio is accumulated when $(1-f/f_0)$ is much larger 
than ${\cal V}$. This is reflected in Eq.~(\ref{5.4}),
below.

\section{Schwarzschild black hole}

To give concreteness to the framework introduced in 
the preceding section, and to prepare the way for the 
next, here we consider the special case in which the 
mass $M$ is a Schwarzschild black hole. In this and the 
following section we assume that general relativity 
gives the correct description of gravitational radiation. 

In this context, the signal's dominant frequency is at 
twice the orbital frequency, so that $2\pi F = 
2\Omega = 2 (M/r^3)^{1/2}$, where $r$ is the orbital
radius in Schwarzschild coordinates. Since the innermost
stable circular orbit is located at $r=6M$ \cite{MTW}, 
Eq.~(\ref{2.2}) gives
\begin{equation}
v_0 = 6^{-1/2} \simeq 0.4083.
\label{3.1}
\end{equation}

The gravitational waves emitted by a mass $\mu$
in circular motion near the ISCO of a Schwarzschild black
hole of mass $M$ can be calculated using the methods of 
black-hole perturbation 
theory \cite{PoissonI,PoissonII,TagoshiNakamura}.
The numerical evaluation of $dE/dt$ then yields estimates for 
the constants $p$ and $\hat{p}$ appearing in Eq.~(\ref{2.3}). 
Using the results presented in Ref.~\cite{PoissonVI}, we obtain
\begin{equation}
p \simeq 7.30, \qquad \hat{p} \simeq 3.78.
\label{3.2}
\end{equation}

On the other hand, $dE/dF$ can be calculated analytically, 
since this quantity relates orbital 
energy and angular velocity for circular geodesics
of the Schwarzschild spacetime. An elementary calculation
yields
\begin{equation}
E = \mu \bigl(1-2v^2\bigr)\bigl(1-3v^2\bigr)^{-1/2},
\label{3.3}
\end{equation}
where $v = (M/r)^{1/2} = (M\Omega)^{1/3} = (\pi M F)^{1/3}$. 
After differentiation,
\begin{equation}
\frac{dE}{dF} = -\frac{\pi \mu M}{3v}\,
\bigl(1-6v^2\bigr)\bigl(1-3v^2\bigr)^{-3/2}.
\label{3.4}
\end{equation}
Finally, expansion about $F=f_0 = (6^{3/2} \pi M)^{-1}$ yields
Eq.~(\ref{2.4}), with
\begin{equation}
q = \frac{4\sqrt{2}\pi}{9} \simeq 1.9746, \qquad
\hat{q} = \frac{1}{2}.
\label{3.5}
\end{equation}

We now use the results obtained thus far to estimate 
the parameters which characterize the gravitational-wave 
signal. 

We begin with the final-frequency parameter, $f_0$. Use
of Eqs.~(\ref{2.2}) and (\ref{3.1}) yields the relationship
between $f_0$ and $M$, which we write as
\begin{equation}
f_0 = 4.40\, \biggl(\frac{v_0}{0.4083} \biggr)^3\, 
\biggl(\frac{M}{10^6\, M_\odot} \biggr)^{-1}\ 
\mbox{mHz}.
\label{3.6}
\end{equation}
We have chosen a fiducial value of $1 \times
10^6\ M_\odot$ for $M$, to ensure that for such a mass, 
$f_0 = 4.4 \times 10^{-3}\ \mbox{Hz}$, where LISA's 
sensitivity is near maximum.
In the way it is written, Eq.~(\ref{3.6}) remains 
valid even when the mass $M$ is not a Schwarzschild black 
hole; in such a situation $v_0$ would differ from $6^{-1/2}$. 
The same will be true for most of the equations listed below.

Next, we use Eqs.~(\ref{2.6}), (\ref{3.1}), (\ref{3.2}), 
and (\ref{3.5}) to obtain
\begin{eqnarray}
{\cal V} &=& 3.95\times 10^{-6}\, 
\biggl(\frac{p}{7.30}\biggr)
\biggl(\frac{q}{1.9746}\biggr)^{-1}
\biggl(\frac{\mu}{10\, M_\odot}\biggr)
\nonumber \\
& & \mbox{} \times
\biggl(\frac{M}{10^6\, M_\odot}\biggr)^{-1}
\biggl(\frac{v_0}{0.4083} \biggr)^5
\label{3.7}
\end{eqnarray}
and
\begin{equation}
\alpha \simeq 1.62.
\label{3.8}
\end{equation}

Continuing with our estimates, we consider a situation
in which the gravitational waves are monitored
during an entire year before the mass $\mu$
reaches the innermost stable circular orbit. We ask:
what is the corresponding fractional change in
frequency? To answer this we integrate Eq.~(\ref{2.5})
for $t(F)$:
\begin{equation}
t(F) = t_0 - \frac{3}{2\pi {\cal V} f_0}\,
(1 - F/f_0)^2\, (1 + \cdots).
\label{3.9}
\end{equation}
Choosing $f_i$ to be the initial frequency, letting
$\Delta t \equiv t_0 - t(f_i)$, $\varepsilon  \equiv 
1 - f_i/f_0$, and ignoring the corrections terms
in Eq.~(\ref{3.9}), we obtain
\begin{eqnarray}
\varepsilon &=& \biggl( \frac{2\pi}{3}\, 
{\cal V} f_0 \Delta t \biggr)^{1/2}
\nonumber \\
& & \label{3.10} \\
&=& 1.07\, 
\biggl(\frac{f_0}{\mbox{4.40\ mHz}}\biggr)^{1/2}
\biggl(\frac{{\cal V}}{3.95\times 10^{-6}}\biggr)^{1/2}
\biggl(\frac{\Delta t}{\mbox{yr}}\biggr)^{1/2}.
\nonumber
\end{eqnarray}
We recall that according to Eqs.~(\ref{2.12})
and (\ref{3.7}), the signal must be cut off at
a final frequency $f_f$ given by
\begin{equation}
1-f_f/f_0 = {\cal V} \sim 10^{-6}.
\label{3.11}
\end{equation}

Finally, we estimate $\cal N$, the number of wave cycles 
received by the detector during the year's worth of observation. 
Even though the signal is not monochromatic, for
our purposes it is sufficient to {\it define}
\begin{eqnarray}
{\cal N} &=& f_0 \Delta t \nonumber \\
&=& 1.39\times 10^5\, \biggl(\frac{f_0}{\mbox{4.40\ mHz}}\biggr)
\biggl(\frac{\Delta t}{\mbox{yr}}\biggr)
\label{3.12} \\
&=& \frac{3\varepsilon^2}{2\pi{\cal V}}.
\nonumber
\end{eqnarray}
The last line follows from Eq.~(\ref{3.10}) and
is displayed here for future reference. We see from
Eq.~(\ref{3.12}) that for an integration time of
one year, the total number of wave
cycles is quite appreciable.

\section{Kerr black hole}

We now generalize the discussion of the previous section
to the case of a Kerr black hole. 

As before we have that the dominant component of the
gravitational waves has a frequency given by 
$F=\Omega/\pi$. For a Kerr black hole, this 
reads \cite{BPT}
\begin{equation}
\pi M F = \frac{1}{r^{3/2} + \chi}\, .
\label{4.1}
\end{equation}
Here, we use a dimensionless 
radial coordinate $r$ which is a rescaled
version of the usual Boyer-Lindquist coordinate, namely,
$r=r_{\rm BL}/M$. The dimensionless constant $\chi$ is
related to $\bbox{J}$, the intrinsic (spin) angular momentum  
of the black hole, and also to $\bbox{\hat{L}}$, the unit 
vector pointing in the direction of orbital angular momentum: 
\begin{equation}
\chi = \frac{\bbox{J} \cdot \bbox{\hat{L}}}{M^2}.
\label{4.2}
\end{equation}
Thus, $\chi$ is {\it positive} if the orbital motion proceeds 
in the same direction as the black-hole rotation (such an orbit 
will be termed {\it direct}); $\chi$ is {\it negative} if 
the orbital motion proceeds in the direction opposite to 
the black-hole rotation (such an orbit will be termed 
{\it retrograde}). It should be noted that $\chi$ is
restricted to the interval $[-1,1]$; $|\chi| = 1$ 
describes an extreme Kerr black hole.

The innermost stable circular orbit is located at the value of
$r$ which satisfies $r^2 - 6r + 8\chi r^{1/2} - 3\chi^2 = 0$.
The solution is \cite{BPT}
\begin{eqnarray}
\bar{r} &=& 3 + B - \frac{\chi}{|\chi|}\, 
\sqrt{(3-A)(3+A+2B)}, \nonumber \\
A &=& 1 + \bigl(1-\chi^2\bigr)^{1/3} \Bigl[ (1+\chi)^{1/3} +
(1-\chi)^{1/3} \Bigr], \label{4.3} \\
B &=& \sqrt{3\chi^2 + A^2}. \nonumber
\end{eqnarray}
We observe that for $\chi=0$, $\bar{r}=6$,
the Schwarzschild value. For $\chi=-1$ (extreme
black hole, retrograde orbit) we have $\bar{r}=9$, 
while $\bar{r}=1$ for $\chi=1$ (extreme black hole, 
direct orbits). The equation $r=1$ also describes
the event horizon of an extreme Kerr black hole.
It is useful to record the expansion
\begin{eqnarray}
\bar{r} &=& 1 + 2^{2/3}\, (1-\chi)^{1/3} +
\frac{7}{2^{5/3}}\, (1-\chi)^{2/3} 
\nonumber \\ & & \mbox{}+
\frac{15}{16}\, (1-\chi) + \cdots,
\label{4.4}
\end{eqnarray}
which is valid for $\chi$ approaching unity. 

\begin{figure}[t]
\special{hscale=35 vscale=35 hoffset=-20.0 voffset=20.0
         angle=-90.0 psfile=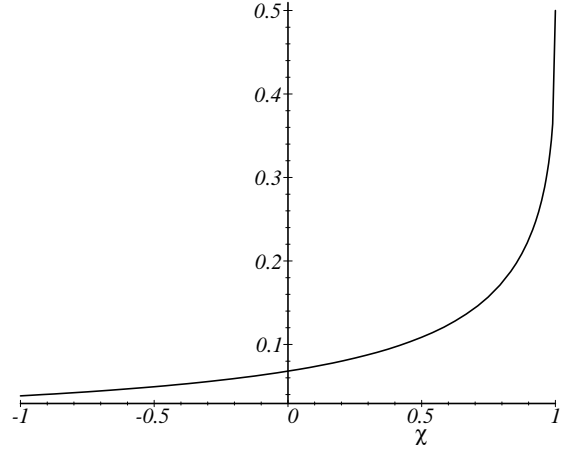}
\vspace*{2.6in}
\caption[Fig. 1]{A plot of $\pi M f_0 \equiv {v_0}^3$ 
as a function of $\chi$.}
\end{figure}

\begin{table}[t]
\caption{The tabulated form of $p$ and $\hat{p}$ as
functions of $\chi$. These values were obtained
from numerical data provided by Masaru Shibata.}
\begin{tabular}{ddd}
$\chi$ & $p$ & $\hat{p}$ \\
\hline
-0.9 & 7.79 & 3.76 \\
-0.8 & 7.75 & 3.77 \\
-0.7 & 7.70 & 3.78 \\
-0.6 & 7.65 & 3.78 \\
-0.5 & 7.60 & 3.78 \\
-0.4 & 7.55 & 3.78 \\
-0.3 & 7.49 & 3.78 \\ 
-0.2 & 7.43 & 3.78 \\
-0.1 & 7.36 & 3.78 \\
 0.0 & 7.30 & 3.78 \\
 0.1 & 7.21 & 3.77 \\
 0.2 & 7.12 & 3.80 \\
 0.3 & 7.02 & 3.77 \\
 0.4 & 6.90 & 3.77 \\
 0.5 & 6.75 & 3.76 \\
 0.6 & 6.57 & 3.75 \\
 0.7 & 6.33 & 3.72 \\
 0.8 & 5.95 & 3.66 \\
 0.9 & 5.22 & 3.46 
\end{tabular}
\end{table}

Combining Eqs.~(\ref{2.2}) and (\ref{4.1}) gives
\begin{equation}
\pi M f_0 \equiv {v_0}^3 = \frac{1}{
{\bar{r}}^{3/2} + \chi}.
\label{4.5}
\end{equation}
A plot of ${v_0}^3(\chi)$ is provided in Fig.~1. We note that
for $\chi=0$, ${v_0}^3$ reduces to the Schwarzschild value,
$6^{-3/2} \simeq 0.0680$. For $\chi=-1$, 
${v_0}^3 = 1/26 \simeq 0.0385$,
while ${v_0}^3 = 1/2$ for $\chi=1$. 
We also calculate, for orbits approaching
the innermost stable circular orbit,
\begin{eqnarray}
1-F/f_0 &=& \frac{r^{3/2} - {\bar{r}}^{3/2}}
{r^{3/2} + \chi} 
\nonumber \\
&=& \frac{3 {\bar{r}}^{1/2}}
{2 \bigl({\bar{r}}^{3/2} + \chi\bigr)}\,
(r-\bar{r}) 
\nonumber \\ & & \mbox{}
-\frac{3\bigl(5{\bar{r}}^{3/2} - \chi\bigr)}
{8{\bar{r}}^{1/2}\bigl({\bar{r}}^{3/2} + \chi\bigr)}\,
(r-\bar{r})^2 + \cdots.
\label{4.6}
\end{eqnarray}
This series can be inverted to give
\begin{eqnarray}
r - \bar{r} &=& 
\frac{2 \bigl({\bar{r}}^{3/2} + \chi\bigr)}
{3\, {\bar{r}}^{1/2}}\, (1-F/f_0) 
\nonumber \\ & & \mbox{} +
\frac{\bigl({5\bar{r}}^{3/2} - \chi\bigr)
\bigl({\bar{r}}^{3/2} + \chi\bigr)}
{9\, {\bar{r}}^2}\, (1-F/f_0)^2  
\nonumber \\ & & \mbox{} + \cdots.
\label{4.7}
\end{eqnarray}
We will use this result in order to reproduce,
for the specific case of a Kerr black hole, 
the expansions (\ref{2.3}) and (\ref{2.4})
for $dE/dt$ and $dE/dF$, respectively.

The computation of the gravitational-wave luminosity, 
$\dot{E}_{\rm GW} \equiv - dE/dt$, for a specified 
circular orbit of the Kerr spacetime, 
is carried out numerically
using the methods of black-hole 
perturbation theory \cite{Shibata2}. 
In such calculations, the value of the radius is 
conveniently used to characterize the orbit. 
By computing $\dot{E}_{\rm GW}(\bar{r})$ and 
$\dot{E}_{\rm GW}^\prime (\bar{r})$,
where a prime denotes differentiation with respect to
$r$, we have access to the parameters $p$ and $\hat{p}$.
The precise relations are
\begin{eqnarray}
p &=& \biggl( \frac{M}{\mu} \biggr)^2\,
\frac{\dot{E}_{\rm GW}(\bar{r})}{{v_0}^{10}},
\nonumber \\
& & \label{4.8} \\
\hat{p} &=& - \frac{2}{3\, {\bar{r}}^{1/2}\, {v_0}^3}\,
\frac{\dot{E}_{\rm GW}^\prime (\bar{r})}
{\dot{E}_{\rm GW}(\bar{r})}. \nonumber
\end{eqnarray}
These parameters are listed in Table I for several
values of $\chi$. The data from which $p$ and
$\hat{p}$ were obtained was kindly provided by Masaru
Shibata (Osaka University, Japan). Because it is 
increasingly difficult to calculate 
$\dot{E}_{\rm GW}$ for increasing values of 
$|\chi|$, such computations were not carried 
out for $|\chi|>0.9$.

As for the case of a Schwarzschild black hole, 
the calculation of $dE/dF$, and its expansion near the
innermost stable circular orbit, proceeds analytically.

We begin with the relation between orbital energy
and radius \cite{BPT}:
\begin{equation}
E/\mu = \frac{r^{3/2} - 2 r^{1/2} + \chi}
{r^{3/4}(r^{3/2} - 3 r^{1/2} + 2\chi)^{1/2}}.
\label{4.9}
\end{equation}
After differentiation and upon using Eq.~(\ref{4.1}),
we obtain
\begin{equation}
\frac{dE}{dF} = -\frac{\pi \mu M}{3}\,
\frac{(r^2-6r + 8\chi r^{1/2}-3\chi^2)
(r^{3/2}+ \chi)^2}{r^{9/4}
(r^{3/2} - 3 r^{1/2} + 2\chi)^{3/2}}.
\label{4.10}
\end{equation}
We recognize, in the numerator, the expression
$r^2-6r + 8\chi r^{1/2}-3\chi^2$ which vanishes when
$r=\bar{r}$. An expansion about this point will
therefore be of the form (\ref{2.4}), as we shall
see presently. 

We point out that this statement is {\it false} in the 
case of direct orbits around an extreme Kerr 
black hole. When $\chi=1$ the denominator of 
$dE/dF$ also goes to zero at $r=\bar{r}=1$, and 
as a consequence, $dE/dF$ does not vanish 
at $r=\bar{r}$. Instead,
\begin{equation}
\lim_{r\to 1} \Biggl( \frac{dE}{dF} \biggm|_{\chi=1}
\Biggr) = - \frac{16 \pi \mu M}{9 \sqrt{3}}.
\label{4.11}
\end{equation}
We may therefore conclude that the condition 
$r=\bar{r}$ does not, in the case
of direct orbits around an extreme Kerr black hole,
correspond to a genuine innermost stably circular orbit,
since $dE/dF$ fails to go to zero there.
The case $\chi=1$ therefore represents a singular
limit in the framework introduced in Sec.~II, and
we shall exclude it from further considerations.
More details about this point can be found in the 
Appendix.

\begin{figure}[t]
\special{hscale=35 vscale=35 hoffset=-20.0 voffset=20.0
         angle=-90.0 psfile=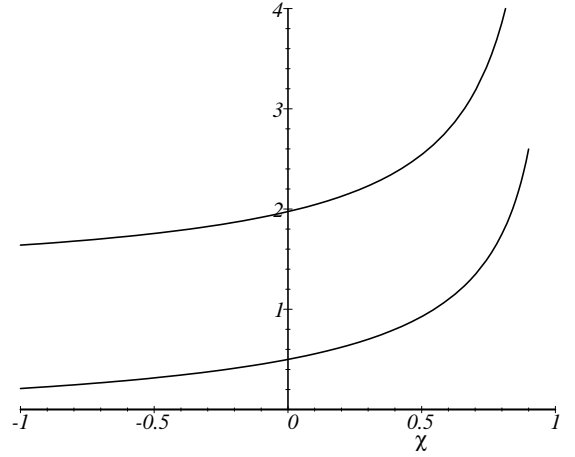}
\vspace*{2.6in}
\caption[Fig. 2]{Plots of $q$ (higher curve) and $\hat{q}$ 
(lower curve) as functions of $\chi$. Both quantities 
diverge as $\chi \to 1$.}
\end{figure}

As was stated before, expansion of Eq.~(\ref{4.10})
about $r=\bar{r}$ and substitution of
Eq.~(\ref{4.7}) reproduces Eq.~(\ref{2.4}),
with
\begin{equation}
q = \frac{4\pi}{9 {v_0}^8 {\bar{r}}^{13/4}\,
\bigl({\bar{r}}^{3/2} - 3 {\bar{r}}^{1/2}
+ 2\chi\bigr)^{1/2}},
\label{4.12}
\end{equation}
and
\begin{equation}
\hat{q} = \frac{-{\bar{r}}^3 + 15 {\bar{r}}^2 +
3\chi {\bar{r}}^{3/2} - 39\chi {\bar{r}}^{1/2}
+ 22\chi^2}{6 {\bar{r}}^{3/2} \bigl({\bar{r}}^{3/2} 
- 3 {\bar{r}}^{1/2} + 2\chi\bigr)}.
\label{4.13}
\end{equation}
Plots of $q(\chi)$ and $\hat{q}(\chi)$ are provided in Fig.~2.
We observe that for $\chi=0$, Eqs.~(\ref{4.12}) and
(\ref{4.13}) reduce to the Schwarzschild expressions given
by Eq.~(\ref{3.5}). For $\chi=-1$,
$q \simeq 1.6401$ and $\hat{q} = 17/81 \simeq 0.2100$. For
$\chi$ approaching $+1$,
$q \sim (16\sqrt{12}\pi/27)\,(1-\chi)^{-1/3}$ and
$\hat{q} \sim (2^{7/3}/3)\,(1-\chi)^{-1/3}$. In view of
Eq.~(\ref{4.11}), it is not surprising that these
quantities should diverge in this limit.  

\begin{figure}[t]
\special{hscale=35 vscale=35 hoffset=-20.0 voffset=20.0
         angle=-90.0 psfile=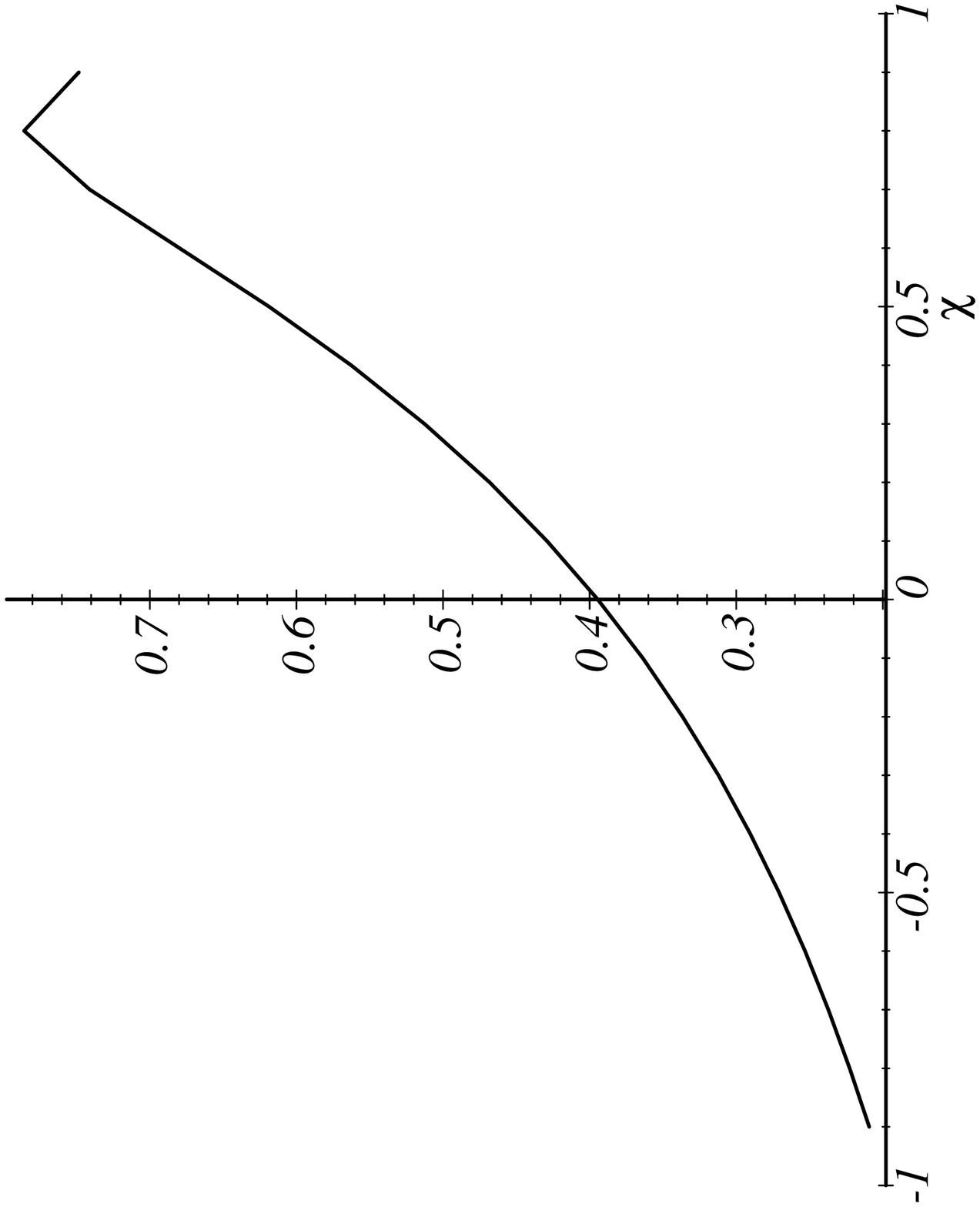}
\vspace*{2.6in}
\caption[Fig. 3]{A plot of $(M/\mu)\, {\cal V}$ 
as a function of $\chi$. The computed data points at
$\chi = \{-0.9, -0.8, \cdots, 0.9\}$ are joined by
straight line segments.}
\end{figure}

\begin{figure}[t]
\special{hscale=35 vscale=35 hoffset=-20.0 voffset=20.0
         angle=-90.0 psfile=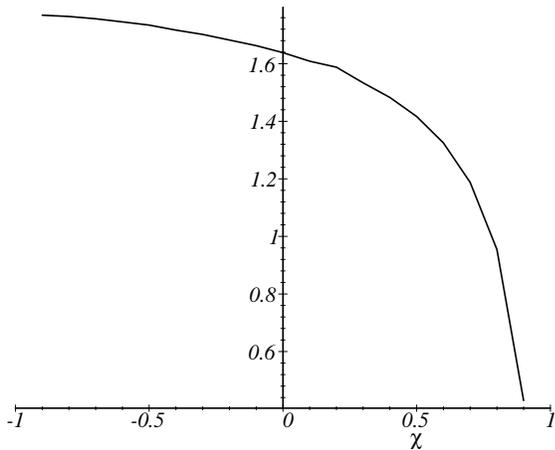}
\vspace*{2.6in}
\caption[Fig. 4]{A plot of $\alpha$ as a function of $\chi$.
The computed data points at $\chi = \{-0.9, -0.8, \cdots, 0.9\}$ 
are joined by straight line segments.}
\end{figure}

We use our results for $v_0$, $p$, $\hat{p}$, $q$, and
$\hat{q}$ to calculate ${\cal V}$ and $\alpha$
according to Eq.~(\ref{2.6}). Plots of these
quantities are provided in Figs.~3 and 4. 

Before leaving this section we point out that
Eqs.~(\ref{3.6}), (\ref{3.7}), and 
(\ref{3.10})--(\ref{3.12}), which provide estimates 
for $f_0$, ${\cal V}$, $f_i/f_0$, $f_f/f_0$,
and $\cal N$, are quite general: they therefore apply 
equally well to the case of a rotating black hole. 

\section{Estimation of signal parameters}

\subsection{Introduction}

The last two sections were devoted to
special cases of the general framework introduced in
Sec.~II. In such cases there exists a specific relationship
between the signal parameters $\{f_0,{\cal V},\alpha\}$
and the source parameters $\{\mu,M,\chi\}$. 
For example, in the case of a Kerr black hole, 
$\alpha$ is a function of the single variable $\chi$;
this function is represented in Fig.~4.

We now return to the general framework of Sec.~II, in
which the nature of the mass $M$ is left
unspecified. The framework also makes no specific choice
regarding the correct description of gravitational
radiation.

In this framework the gravitational-wave signal is 
given by Eq.~(\ref{2.10}), which we rewrite as
\begin{eqnarray}
\tilde{h}(f) &=& {\cal A}\, (1-f/f_0)^{1/2}\,
e^{i \psi(f)},
\nonumber \\
\psi(f) &=& 2\pi f t_0 - \phi_0 + \phi(f),
\label{5.1} \\
\phi(f) &=& \frac{(1-f/f_0)^3}{{\cal V}}\, \bigl[
1 + \alpha (1-f/f_0)\bigr].
\nonumber
\end{eqnarray}
For the purpose of this section we consider 
Eq.~(\ref{5.1}) to be exact, ignoring all relative 
corrections of order $(1-f/f_0)$ in the amplitude, and all
relative corrections of order $(1-f/f_0)^2$ in the phase. 

The signal is characterized by six parameters,
which we group into the vector
\begin{equation}
\bbox{\theta} = (\ln{\cal A}, \ln f_0, \ln {\cal V}, 
\alpha, t_0, \phi_0).
\label{5.2}
\end{equation}
The purpose of this section is to calculate, for a
given signal strength, how accurately the parameters
$\bbox{\theta}$ can be estimated during a 
gravitational-wave measurement.

\subsection{Parameter estimation}

The statistical theory of signal detection and 
measurement \cite{Helstrom,Wainstein} has repeatedly 
been applied to the specific case of gravitational 
waves \cite{Thorne,Schutz}. 
For a complete exposition, we refer the reader to 
Refs.~\cite{Finn,CutlerFlanagan,PoissonWill}.
That the theory must be statistical in character follows 
from the facts that gravitational-wave detectors are noisy, and 
that the noise is a random process, here assumed to be stationary 
and Gaussian. In this subsection we apply the theory to calculate, 
in the limit of large signal-to-noise ratio, the expected statistical 
errors associated with the estimation of the signal parameters 
$\bbox{\theta}$.

We first introduce the inner product $(\cdot | \cdot)$: 
If $a$ and $b$ are two functions of time, with
Fourier transforms $\tilde{a}$ and $\tilde{b}$, then
\begin{equation}
(a|b) = 2 \int_0^\infty 
\frac{\tilde{a}^*(f)\, \tilde{b}(f) + 
\tilde{a}(f)\, \tilde{b}^*(f)}{S_n(f)}\, df.
\label{5.3}
\end{equation}
Here, an asterisk denotes complex conjugation, and $S_n(f)$ is
the spectral density of the detector noise, the Fourier transform
of the noise's autocorrelation function. The detector has good
sensitivity where $S_n(f)$ is small, and poor sensitivity where 
$S_n(f)$ is large. For LISA, $S_n(f)$ is small in the frequency 
bandwidth between $10^{-4}$ and $10^{-1}$ Hz \cite{LISA}. 
The precise behavior, and overall normalization, of $S_n(f)$ 
will be of no concern to us.

The {\it signal-to-noise ratio} $\rho$ associated with the
measurement of a signal $h(t)$ by a detector with spectral 
density $S_n(f)$ is given by $\rho^2 = (h|h)$. This is simply 
evaluated for our model signal (\ref{5.1}). Expanding the 
spectral density as $S_n(f) = S_n(f_0)[1 + O(1-f/f_0)]$, 
neglecting the second term for consistency, and integrating 
from $f=f_i$ to $f=f_f$, we obtain
\begin{equation}
\rho^2 = \frac{2 f_0 |{\cal A}|^2 \varepsilon^2}{S_n(f_0)}.
\label{5.4}
\end{equation}
Here, $\varepsilon = 1-f_i/f_0$, and is given by Eq.~(\ref{3.10}). 
We have neglected ${\cal V}^{2}$ in front of $\varepsilon^2$,
cf.\ Eq.~(\ref{3.11}). From Eq.~(\ref{5.4}) we see that the
signal-to-noise ratio is proportional to the signal's
amplitude times the square root of its duration 
(as measured by $\varepsilon$), and inversely 
proportional to the noise's amplitude.

We calculate the anticipated accuracy with which the parameters 
$\bbox{\theta}$ can be estimated during a gravitational-wave
measurement. In the limit of large signal-to-noise ratio (or
small statistical errors), the
calculation involves computing, and then inverting, the Fisher
information matrix $\Gamma_{ab} = ( h_{,a}| h_{,b})$,
where $h_{,a} = \partial h / \partial \theta^a$ is the
partial derivative of the signal with respect to the
parameter $\theta^a$. 

With the ordering provided by Eq.~(\ref{5.2}), the
partial derivatives are given by
\begin{equation}
\tilde{h}_{,a} = g_a\, \tilde{h},
\label{5.5}
\end{equation}
where
\begin{eqnarray}
g_1 &=& 1, \nonumber \\
g_2 &=& \frac{1-x}{2x} 
\biggl[ 1 + \frac{2 i x^3}{{\cal V}}\, (3+4\alpha x) \biggr],
\nonumber \\
g_3 &=& - \frac{i x^3}{{\cal V}}\, (1+\alpha x),
\nonumber \\
g_4 &=& \frac{i x^4}{{\cal V}}, \nonumber \\
g_5 &=& i (2\pi f_0)(1-x), \nonumber \\
g_6 &=& -i, 
\label{5.6}
\end{eqnarray}
and $x=1-f/f_0$. Then, using Eq.~(\ref{5.4}) and 
the same approximation for $S_n(f)$ as was used before, 
we have that
\begin{equation}
\Gamma_{ab} = \frac{\rho^2}{\varepsilon^2}\,
\int_{{\cal V}}^\varepsilon \bigl[ g_a^*(x)\, g_b(x) +
g_a(x)\, g_b^*(x) \bigr] x \, dx.
\label{5.7}
\end{equation}
These integrations are easily carried out. Because
its expression is rather large, the Fisher matrix 
will not be displayed here.

The meaning of the Fisher matrix is as 
follows \cite{Finn,CutlerFlanagan}. Because of random 
noise, the output of a gravitational-wave detector is also random, 
even when a signal of the form $h(t;\bbox{\theta})$ is known to be 
present. Consequently, an estimation of the parameters $\bbox{\theta}$ 
comes with inevitable statistical uncertainty. It can be shown that
in the limit of large signal-to-noise ratio, the probability 
distribution function for the parameters $\bbox{\theta}$ is of the 
Gaussian form $\exp[-\frac{1}{2} \Gamma_{ab} (\theta^a-\bar{\theta}^a)
(\theta^b-\bar{\theta}^b)]$, where $\bbox{\bar{\theta}}$ denotes
the true value of the signal parameters. 
It follows that the variance-covariance
matrix, $\Sigma^{ab} \equiv \langle (\theta^a-\bar{\theta}^a)
(\theta^b-\bar{\theta}^b) \rangle$, where $\langle \cdot \rangle$
denotes an averaging with respect to the specified distribution, 
is given simply by the inverse of the Fisher matrix,
\begin{equation}
\Sigma^{ab} = (\bbox{\Gamma}^{-1})^{ab}.
\label{5.8}
\end{equation}
In terms of this, the ``one-sigma'' statistical error associated 
with the estimation of parameter $\theta^a$ is given by
\begin{equation}
\Delta \theta^a = \sqrt{\Sigma^{aa}},
\label{5.9}
\end{equation}
where there is no summation over the repeated indices.

To invert the Fisher matrix we proceed as follows. After
computing the matrix we use Eq.~(\ref{3.12}) to express 
${\cal V}$ in terms of $\varepsilon$ and $\cal N$. 
The matrix is then inverted 
algebraically, using the symbolic manipulator Maple. 
Thus far all expressions are exact, and quite large.
To simplify, we take advantage of the fact that according 
to Eq.~(\ref{3.12}), $\cal N$ is numerically very large.
We therefore expand our exact expression for $\bbox{\Sigma}$
in powers of $1/{\cal N}$, keeping only the leading-order term. 
It can be verified that the resulting matrix is equal to 
$\bbox{Sigma}$, apart from a fractional correction 
of order $1/{\cal N}^2$. 

We obtain the following results:
\begin{eqnarray}
\frac{\Delta {\cal A}}{\cal A} &=& \frac{1}{\rho},
\nonumber \\
\frac{\Delta f_0}{f_0} &=& \frac{280\sqrt{3}}{2\pi 
{\cal N} \rho}, \nonumber \\
\frac{\Delta {\cal V}}{{\cal V}} &=&
\frac{1890 \sqrt{a}}
{2\pi {\cal N} \varepsilon \rho},
\nonumber \\
\Delta \alpha &=& \frac{378\sqrt{5 b}}
{2\pi {\cal N} \varepsilon^2 \rho}, 
\nonumber \\
\Delta t_0 &=& \frac{105\sqrt{2}}{2\pi f_0
\varepsilon \rho}, 
\nonumber \\
\Delta \phi_0 &=& \frac{105\sqrt{2 c}}
{\varepsilon \rho},
\label{5.10.5}
\end{eqnarray}
where
\begin{eqnarray}
a(\varepsilon,\alpha) &=& 1 - 
\biggl(\frac{32}{21} - \frac{128}{63} 
\alpha \biggr)\, \varepsilon 
\nonumber \\ & & \mbox{}
+ \biggl(\frac{16}{27} - \frac{128}{81}\, \alpha + 
\frac{256}{243}\, \alpha^2 \biggr)\, \varepsilon^2,
\nonumber \\
b(\varepsilon,\alpha) &=& 1 + 
\frac{40}{9}\, \alpha\, \varepsilon + 
\frac{5}{9}\, (2+17\alpha) \alpha\, \varepsilon^2 
\nonumber \\ & & \mbox{} +
\frac{160}{63}\, (1+4\alpha) \alpha^2\, \varepsilon^3 +
\frac{80}{243}\, (1+4\alpha)^2 \alpha^2\, \varepsilon^4,
\nonumber \\
c(\varepsilon) &=& 1 - 
\frac{4}{21}\, \varepsilon + \frac{1}{98}\, \varepsilon^2.
\label{5.10.6}
\end{eqnarray}

To put this in a more concrete form we evaluate
$a$, $b$, and $c$ at $\varepsilon = 1.1$ and $\alpha = 1.6$; 
cf.\ Eqs.~(\ref{3.8}) and (\ref{3.10}). This gives
$a\simeq 3.8$, $b\simeq 170$, and $c\simeq 0.80$. We
then choose ${\cal N} = 1.4 \times 10^5$ and $f_0 =
4.4\ \mbox{mHz}$ as fiducial values. Substituting all
this, we obtain
\begin{eqnarray}
\frac{\Delta {\cal A}}{\cal A} &=& \frac{1}{\rho},
\nonumber \\
\frac{\Delta f_0}{f_0} &\simeq& 
\frac{5.5 \times 10^{-4}}{\rho}\,
\biggl( \frac{{\cal N}}{1.4 \times 10^5} \biggr)^{-1},
\nonumber \\
\frac{\Delta {\cal V}}{{\cal V}} &\simeq&
\frac{3.8 \times 10^{-3}}{\rho}\,
\biggl( \frac{a}{3.8} \biggr)^{1/2} 
\biggl( \frac{\varepsilon}{1.1} \biggr)^{-1}
\biggl( \frac{{\cal N}}{1.4 \times 10^5} \biggr)^{-1},
\nonumber \\
\Delta \alpha &\simeq& 
\frac{1.0 \times 10^{-2}}{\rho}\,
\biggl( \frac{b}{170} \biggr)^{1/2} 
\biggl( \frac{\varepsilon}{1.1} \biggr)^{-2}
\biggl( \frac{{\cal N}}{1.4 \times 10^5} \biggr)^{-1},
\nonumber \\
\Delta t_0 &\simeq& 
\frac{81\ \mbox{min}}{\rho}\,
\biggl( \frac{f_0}{4.4\ \mbox{mHz}} \biggr)^{-1}
\biggl( \frac{\varepsilon}{1.1} \biggr)^{-1},
\nonumber \\
\Delta \phi_0 &\simeq& 
\frac{19\ \mbox{cycles}}{\rho}\,
\biggl( \frac{c}{0.80} \biggr)^{1/2} 
\biggl( \frac{\varepsilon}{1.1} \biggr)^{-1}.
\label{5.11}
\end{eqnarray}

From these expressions we observe that: (i) The 
relative accuracy with which ${\cal A}$, the amplitude 
parameter, can be estimated is precisely equal to the 
inverse signal-to-noise ratio. We shall call this level of
accuracy the ``signal-to-noise ratio level''. (ii) The 
estimation of $f_0$ benefits the most from the 
large number of wave cycles; its accuracy is typically 
three orders of magnitude better than the signal-to-noise
ratio level. (iii) The estimation of ${\cal V}$ and 
$\alpha$ also benefits from the large value of $\cal N$;
their accuracies are respectively three and two orders
of magnitude better than the signal-to-noise ratio 
level. (iv) The uncertainties associated with the
parameters $t_0$ and $\phi_0$ are independent
of $\cal N$.

\subsection{Measurement of black-hole parameters}

For the purpose of this subsection we return to the
special case in which the mass $M$ is a Kerr black 
hole (Sec.~IV). The question we ask is: What does a 
measurement of the signal parameters
$\{f_0,{\cal V},\alpha\}$ tell us about the
source parameters $\{\mu,M,\chi\}$? 

Inspection of Figs.~1, 3, and 4 reveals that
the quantities $\pi M f_0 \equiv {v_0}^3$, 
$\eta^{-1} {\cal V}$, and $\alpha$ are all 
functions of the single variable $\chi$. 
Here we have defined $\eta$ to be the mass ratio:
\begin{equation}
\eta = \frac{\mu}{M}.
\label{5.12}
\end{equation}
We therefore write
\begin{eqnarray}
\pi M f_0 &=& A(\chi),
\nonumber \\
\eta^{-1} {\cal V} &=& B(\chi),
\label{5.13} \\
\alpha &=& C(\chi).
\nonumber
\end{eqnarray}
It follows from these relations that a measurement of
$\alpha$ yields the value of $\chi$. Knowing $\chi$,
a measurement of ${\cal V}$ then gives $\eta$. And
finally, a measurement of $f_0$ returns the value of $M$. 
In this way, the source parameters can all be estimated 
from a measurement of the gravitational-wave signal. The 
question remains as to how accurately all this can be done.
This is the question we turn to next.

We first consider the estimation of $\chi$, the black-hole
rotation parameter. It follows from Eq.~(\ref{5.13}) that
in the limit of small measurement uncertainties, the error 
in $\chi$ is given by
\begin{equation}
\Delta \chi = \frac{1}{|C'(\chi)|}\, \Delta \alpha,
\label{5.14}
\end{equation}
where a prime denotes differentiation with respect to
$\chi$. To express this in a more concrete form it would
be desirable to explicitly compute the function $C'(\chi)$.
Unfortunately, this cannot be done reliably, since $C(\chi)$
was evaluated at only a few widely separated
points. Nevertheless, we crudely estimate $C'(\chi)$ to
be $0.2$ at the somewhat representative point $\chi=0$. 
This yields
\begin{equation}
\Delta \chi \simeq 5.0\, \Delta \alpha
\qquad \mbox{for $\chi=0$}.
\label{5.15}
\end{equation}
It can be seen from Fig.~4 that the uncertainty in $\chi$ 
would be larger than this for negative values of $\chi$,
while it would be smaller (and going to zero for $\chi 
\to 1$) for positive values. Using Eq.~(\ref{5.11}), 
Eq.~(\ref{5.15}) means that for typical situations, 
$\Delta \chi \simeq 5.0 \times 10^{-2}/\rho$. The 
rotation parameter can therefore be estimated with a
degree of accuracy approximately two orders of
magnitude better than the signal-to-noise ratio level.

We now repeat this procedure for $\eta$, the
mass-ratio parameter. In the limit of small 
uncertainties, Eq.~(\ref{5.13}) implies
\begin{equation}
\frac{\Delta \eta}{\eta} = \biggl| \frac{B'(\chi)}{B(\chi)}
\biggr| \Delta \chi + \frac{\Delta {\cal V}}{{\cal V}}.
\label{5.16}
\end{equation}
A crude estimate reveals that $|B'/B| \simeq 1.1$ at
$\chi=0$. This means that the uncertainty in $\chi$
dominates the uncertainty in $\eta$, and we find
\begin{equation}
\frac{\Delta \eta}{\eta} \simeq 1.1 \Delta \chi
\qquad \mbox{for $\chi=0$}.
\label{5.17}
\end{equation}
For typical situations, $\Delta \eta/\eta \simeq
5.5 \times 10^{-2}/\rho$, a degree of accuracy one
order of magnitude better than the signal-to-noise 
ratio level.

\begin{figure}[t]
\special{hscale=35 vscale=35 hoffset=-20.0 voffset=20.0
         angle=-90.0 psfile=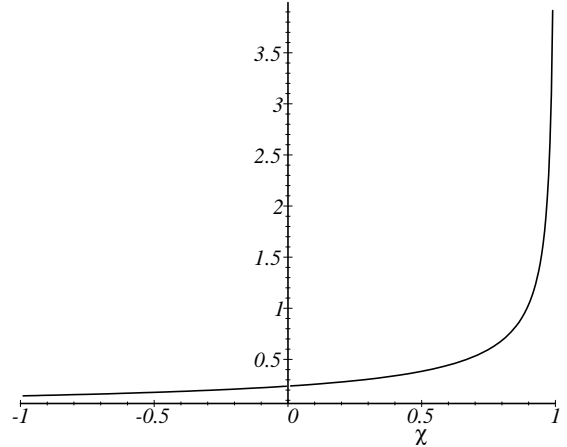}
\vspace*{2.6in}
\caption[Fig. 5]{A plot of $A'/\pi A$ as a function of
$\chi$.}
\end{figure}

Finally, we consider the accuracy with which
$M$, the black-hole mass, can be estimated. 
We once more use Eq.~(\ref{5.13}) to derive
\begin{equation}
\pi \frac{\Delta M}{M} = \biggl| \frac{A'(\chi)}{A(\chi)}
\biggr| \Delta \chi + \frac{\Delta f_0}{f_0}.
\label{5.18}
\end{equation}
A plot of $A'/\pi A$ is provided in Fig.~5. Once more
we find that the uncertainty in $\chi$ dominates over
the other contribution. Evaluation at $\chi=0$ yields
\begin{equation}
\frac{\Delta M}{M} \simeq 0.238 \Delta \chi
\qquad \mbox{for $\chi=0$}.
\label{5.19}
\end{equation}
It can be seen from Fig.~5 that the uncertainty in $M$
would be smaller than this for negative values of $\chi$, 
larger for positive values, and significantly larger for
$\chi$ approaching unity. For typical situations, 
$\Delta M/M \simeq 2.4 \times 10^{-3}/\rho$, 
a degree of accuracy three orders of magnitude better
than the signal-to-noise ratio level.

The bottom line is that during a 
gravitational-wave measurement, the source parameters 
$\{\mu,M,\chi\}$ can be estimated with a degree of accuracy 
that is much better than the signal-to-noise ratio level.

\subsection{Testing general relativity}

Since the framework of Sec.~II makes 
no explicit reference to a particular
model for the mass $M$, nor to a particular theory
of gravitation, we may ask whether it leads to a
viable strategy for testing the strong-field predictions of general 
relativity. More precisely, we ask: Can a measurement of the 
signal parameters $\{f_0, {\cal V}, \alpha \}$ allow us to
test the hypothesis according to which the central object
is a Kerr black hole and the correct theory of gravitation 
is general relativity?

Clearly, measuring $f_0$ and ${\cal V}$ if of no use in 
this purpose. This is because $f_0$ and ${\cal V}$
respectively depend on $M$ and $\eta = \mu/M$, 
which are not known prior to the measurement. 
Therefore, measuring $f_0$ and ${\cal V}$ offers no
way of distinguishing two different models for the source
which, by adjusting the values of $M$ and $\mu$, give
identical values for $f_0$ and ${\cal V}$.

The parameter $\alpha$ is more useful. It is a 
consequence of the standard model (the mass $M$ is
a Kerr black hole and the correct theory of 
gravitation is general relativity) that $\alpha$ is restricted
to the interval $I_{\rm sm} = (-\infty, 1.8)$. Measuring $\alpha$ 
therefore leads to a way of testing general relativity, since
a measured value of $\alpha$ outside $I_{\rm sm}$ would 
unambiguously invalidate the standard model, 

Of course, while viable in principle, this test is very much a 
partial one, and leaves much to be desired. For example, it is 
impossible to say which aspect of the standard model is 
violated when $\alpha \not\in I_{\rm sm}$. Another shortcoming 
is that there could well exist a wide 
class of alternative models for which $\alpha$ is restricted 
to an interval $I$ which is a {\it subset} of $I_{\rm sm}$. 
The test offers no way of ruling out models belonging 
to this class. Finally, we must also recall that the entire 
framework is based upon a rather restrictive assumption: 
that the orbit must be circular and within the equatorial plane.

\section*{Acknowledgments}

I am most grateful to Clifford Will for his constant encouragement 
during the realization of this work, which originated during 
discussions with him. This paper could not have been written without
the kind help of Masaru Shibata, who provided the numerical data used
to construct Table I.  A conversation with Kip Thorne 
was also greatly appreciated. This work was supported by
the Natural Sciences and Engineering Research Council
of Canada, by the National Science Foundation
under Grant No.~PHY 92-22902, and by the
National Aeronautics and Space Administration under
Grant No.~NAGW 3874. 
 
\appendix

\section*{}

In this Appendix we consider the motion of a
test mass in a spacetime satisfying the following
properties: it is stationary and asymptotically flat, 
possesses axial symmetry, and is reflection symmetric 
about the equatorial plane.
We do not assume that the spacetime metric satisfies
the Einstein field equations. However, for simplicity 
we shall assume that the test mass is moving freely, so that 
it follows a geodesic of the spacetime. Generalization
to the presence of external forces (which also satisfy
the symmetry requirements) would be straightforward.
We shall focus our attention on equatorial, circular orbits, 
the meaning of which will be made precise below. 

The results obtained in this Appendix form the basis 
for the construction of the model signal of Sec.~II.

\subsection{Equations of motion}

We introduce coordinates $(t,r,\theta,\phi)$ to chart
the spacetime. We take $t$ to be the time coordinate 
associated with the timelike Killing vector $\bbox{\xi}
= \partial/\partial t$, and normalized 
such that it is equal to proper 
time for observers at rest at infinity. We take $r$ to be 
a well-behaved radial coordinate, such that circular orbits 
(defined in a coordinate-invariant way below)
are described by $r=\mbox{constant}$. We take $\theta$ 
and $\phi$ to be the usual spherical coordinates, such that $\phi$ 
is associated with the Killing vector $\bbox{\chi} = 
\partial / \partial \phi$. In these coordinates, according 
to the specified symmetries, the only off-diagonal component of 
the metric is $g_{t\phi}$. For the specific case of a Kerr
black hole, the coordinates employed reduce to the Boyer-Lindquist
coordinates \cite{MTW}. 

Reflection symmetry about the equatorial plane ensures that 
the motion of a test mass can be restricted to lie in
this plane. (Of course, such a restriction represents a loss
of generality). We therefore take $\theta=\pi/2$, and the 
four-velocity is $u^\alpha = dx^\alpha/d\tau = (\dot{t},
\dot{r}, 0, \dot{\phi})$, where an overdot denotes differentiation
with respect to proper time $\tau$. The quantities
\begin{equation}
\tilde{E} = -\bbox{u} \cdot \bbox{\xi}, \qquad
\tilde{L} = \bbox{u} \cdot \bbox{\chi}
\label{A1}
\end{equation}
are constants of the motion, respectively representing
orbital energy per unit rest mass, and orbital angular
momentum per unit rest mass. 

It is easy to show that Eqs.~(\ref{A1}) imply
\begin{eqnarray}
\dot{t} &=& T(r,\tilde{E},\tilde{L}) \equiv 
\bigl(g_{\phi\phi} \tilde{E} + g_{t\phi} \tilde{L}\bigr)/D,
\nonumber \\
& & \label{A2} \\
\dot{\phi} &=& \Phi(r,\tilde{E},\tilde{L}) \equiv 
-\bigl(g_{t\phi} \tilde{E} + g_{tt} \tilde{L}\bigr)/D,
\nonumber
\end{eqnarray}
where $D = -g_{tt} g_{\phi\phi} + {g_{t\phi}}^2$. On the
other hand, the normalization condition $\bbox{u} \cdot
\bbox{u} = -1$ gives
\begin{equation}
\dot{r}^2 = R(r,\tilde{E},\tilde{L}) = 
-\frac{1}{g_{rr}} + \frac{g_{\phi\phi} \tilde{E}^2
+ 2 g_{t\phi} \tilde{E} \tilde{L} + g_{tt} \tilde{L}^2}{g_{rr} D}.
\label{A3}
\end{equation}
As was indicated, the metric functions are considered
to be functions of $r$ only, since they are evaluated at
$\theta=\pi/2$. 

We define a function $\Omega$ by
\begin{equation}
\Omega(r,\tilde{E},\tilde{L}) = \frac{d\phi}{dt} = 
\frac{\Phi(r,\tilde{E},\tilde{L})}
{T(r,\tilde{E},\tilde{L})}.
\label{A4}
\end{equation}
This represents the angular velocity of the test mass,
as measured by an observer at rest at infinity.

\subsection{Circular orbits}

We define an orbit to be circular if $r$ is constant
along the orbit. The conditions for this to be true are
$\dot{r} = \ddot{r} = 0$. According to Eq.~(\ref{A3}), 
this means 
\begin{equation}
R(r,\tilde{E},\tilde{L}) = \frac{\partial R}{\partial r}
(r,\tilde{E},\tilde{L}) =0.
\label{A5}
\end{equation}
These equations determine the values that $\tilde{E}$ 
and ${\tilde L}$
must take in order to produce a circular orbit at
radius $r$. For an orbit at $r=r_0$, we shall denote 
these values by $\tilde{E}_0$ and $\tilde{L}_0$. 

Let $\Omega_0$ be the angular-velocity function 
evaluated at $r=r_0$, $\tilde{E}=\tilde{E}_0(r_0)$, and 
$\tilde{L}=\tilde{L}_0(r_0)$. We therefore
have the important result that the angular velocity
is uniform along a circular orbit. This provides a 
coordinate-invariant definition of circularity, a 
property that has nothing to do with the geometric
appearance of the orbit. As a condition on our 
choice of radial coordinate, we require that 
$\Omega_0$ be a smooth, monotonic function of $r_0$.  

\subsection{Neighboring circular orbits}

We now consider two neighboring circular orbits, the first
at $r=r_0$ (with associated parameters 
$\tilde{E}_0$ and $\tilde{L}_0$), the second at 
$r=r_1$ (with associated parameters $\tilde{E}_1$ and 
$\tilde{L}_1$). We assume that $r_1 = r_0 + \delta r_0$, 
with $\delta r_0$ a small quantity. Correspondingly,
$\tilde{E}_1 = \tilde{E}_0 + \delta \tilde{E}_0$ and 
$\tilde{L}_1 = \tilde{L}_0 + \delta \tilde{L}_0$.
We want to calculate various relationships between
the displacements $\delta r_0$, $\delta \tilde{E}_0$, and
$\delta \tilde{L}_0$.

From Eq.~(\ref{A5}) we must have 
$R(r_1,\tilde{E}_1,\tilde{L}_1) = 0$.
Expanding this about 
$(r_0,\tilde{E}_0,\tilde{L}_0)$, up to first order
in the displacements, gives
\[
\frac{\partial R}{\partial \tilde{E}}\biggr|_0 \delta \tilde{E}_0 
+ \frac{\partial R}{\partial \tilde{L}}\biggr|_0 \delta \tilde{L}_0 
= 0.
\]
This equation takes a more familiar form if we
use Eq.~(\ref{A3}) to evaluate the partial derivatives.
We quickly arrive at
\begin{equation}
\delta \tilde{E}_0 = \Omega_0\, \delta \tilde{L}_0.
\label{A6}
\end{equation}

Equation (\ref{A5}) also implies that 
$\partial_r R(r_1,\tilde{E}_1,\tilde{L}_1)
=0$. Expanding, using Eq.~(\ref{A3}) to evaluate the
partial derivatives, and Eqs.~(\ref{A4}) and (\ref{A6}) 
to simplify the result, we arrive at
\begin{equation}
\frac{\partial^2 R}{\partial r^2} \biggr|_0\, \delta r_0 = 
\frac{2}{\Omega_0\, g_{rr}}\, \biggl(
\frac{\partial \Phi}{\partial r} - \Omega_0
\frac{\partial T}{\partial r} \biggr) \biggr|_0 \,
\delta {\tilde E}_0.
\label{A7}
\end{equation}
We notice that $(\partial_r \Phi - \Omega \partial_r T)/T =
\partial_r \Omega$. Regularity of the radial coordinate
therefore implies that the right-hand side of Eq.~(\ref{A7}) 
is nonzero whenever $T\neq 0$, or $g_{rr} \neq \infty$.

A similar calculation would yield $\delta\Omega_0$ in terms
of either one of $\delta r_0$, $\delta \tilde{E}_0$, or 
$\delta \tilde{L}_0$.

\subsection{Innermost stable circular orbit}

We assume that the spacetime is such that stable circular
orbits can only exist at radii larger than some limiting value
$r_{\rm isco}$. At $r=r_{\rm isco}$, circular orbits become
unstable, in the following sense: an infinitesimal change 
in the orbital energy (or orbital
angular momentum) produces a finite change in the orbital
radius. The innermost stable circular orbit is therefore
defined to be that particular circular orbit at which
\begin{equation}
\frac{\delta \tilde{E}_0}{\delta r_0} 
\biggr|_{\rm isco} = 0.
\label{A8}
\end{equation}

We shall assume that the condition
\begin{equation}
\frac{2}{\Omega_0 g_{rr}}\, \biggl(
\frac{\partial \Phi}{\partial r} - \Omega_0
\frac{\partial T}{\partial r} \biggr) 
\biggr|_{\rm isco} \neq 0
\label{A9}
\end{equation}
holds. [For the specific case of a non-extreme Kerr
black hole, this condition can be shown to hold for both 
direct and retrograde orbits, for any value of the hole's 
angular momentum. However, for the special case of an 
{\it extreme} Kerr black hole, the condition is violated 
for direct orbits. This is because $g_{rr}|_{\rm isco}
= \infty$, which signals the presence of an event horizon
at $r=r_{\rm isco}$. As a result, $\delta \tilde{E} /
\delta r_0$ {\it never} goes to zero for direct circular 
orbits around an extreme Kerr black hole.]

Combining Eqs.~(\ref{A8}) and (\ref{A9}), we obtain
\begin{equation}
\frac{\partial^2 R}{\partial r^2} \biggr|_{\rm isco} = 0.
\label{A10}
\end{equation}
The innermost stable circular orbit corresponds to an
inflection point of the effective potential $R$. 

We have required that $\Omega_0$ be a smooth, monotonic
function of $r_0$. Equation (\ref{A8}) therefore also
implies that 
\begin{equation}
\frac{\delta \tilde{E}_0}{\delta \Omega_0} \biggr|_{\rm isco} 
= 0.
\label{A11}
\end{equation}
This is a coordinate-invariant characterization of the
innermost stable circular orbit. 

It is easy to check that near $r_0 = r_{\rm isco}$, 
$\delta \tilde{E}_0/\delta \Omega_0 
\propto \partial^2 R/\partial r^2 |_0$. 
We assume that $\partial^3 R/\partial
r^3$ does not vanish at $r_0=r_{\rm isco}$, so that
the right-hand side vanishes linearly with $r_0 - r_{\rm isco}$.
We then quickly arrive at the following estimate:
\begin{equation}
\frac{\delta \tilde{E}_0}{\delta \Omega_0} \sim k
(\Omega_0 - \Omega_{\rm isco}),
\label{A12}
\end{equation}
for some constant $k$. This result provides the
justification for Eq.~(\ref{2.4}), above.


\begin{references}
\bibitem[*]{email} E-mail address: poisson@terra.physics.uoguelph.ca.
\bibitem[\dagger]{pa} Permanent address.
\bibitem{LISA} P. Bender, I. Ciufolini, K. Danzmann, W. Folkner,
        J. Hough, D. Robertson, A. R\"udiger, M. Sandford, 
        R. Schilling, B. Schutz, R. Stebbins, T. Summer,
        P. Touboul, S. Vitale, H. Ward, and W. Winkler,
        {\it LISA: Laser Interferometer Space Antenna for
        the detection and observation of gravitational
        waves}, Pre-Phase S Report, December 1995 (unpublished).
\bibitem{LIGO} A. Abramovici, W.E. Althouse, R.W.P. Drever,
        Y. G{\" u}rsel, S. Kawamura, F.J. Raab, D. Shoemaker,
        L. Siewers, R.E. Spero, K.S. Thorne, R.E. Vogt,
        R. Weiss, S.E. Whitcomb, and M.E. Zucker,
        Science {\bf 256}, 325 (1992). 
\bibitem{CutlerLindblom} C. Cutler and L. Lindblom, {\it
        Gravitational helioseismology?}, Phys. Rev. D (to
        be published).
\bibitem{Hilsetal} D. Hills, P.L. Bender, and R.F. Webbink,
        Astrophys. J. {\bf 360}, 75 (1990).
\bibitem{CaldwellAllen} R.R. Caldwell and B. Allen, Phys. Rev
        D {\bf 45}, 3447 (1992).
\bibitem{HilsBender} D. Hils and P.L. Bender, Astrophys.
        J. {\bf 445}, L7 (1995).
\bibitem{Shibata} M. Shibata, Phys. Rev. D {\bf 50},
        6297 (1994).
\bibitem{Tanakaetal} T. Tanaka, M. Shibata, M. Sasaki,
        H. Tagashi, and T. Nakamura, Prog. Theor. Phys. 
        {\bf 90}, 65 (1993).
\bibitem{CKP} C. Cutler, D. Kennefick, and E. Poisson,
        Phys. Rev. D {\bf 50}, 3816 (1994).
\bibitem{HilsBender2} D. Hils and P.L. Bender (unpublished).
\pagebreak
\bibitem{RyanMM} F.D. Ryan, Phys. Rev. D {\bf 52}, 5707 (1995).
\bibitem{FinnOriThorne} L.S. Finn, A. Ori, and K.S. Thorne
        (unpublished).
\bibitem{PoissonI} E. Poisson, Phys. Rev. D {\bf
        47}, 1497 (1993).
\bibitem{Echeverria} F. Echeverria, Phys. Rev. D 
        {\bf 40}, 3194 (1989). 
\bibitem{Finn} L.S. Finn, Phys. Rev. D {\bf 46}, 5236 (1992).
\bibitem{QNM} S. Chandrasekhar and S.L. Detweiler, Proc. R.
        Soc. London {\bf A344}, 441 (1975).
\bibitem{FinnChernoff} L.S. Finn and D.F. Chernoff
        Phys. Rev. D {\bf 47}, 2198 (1993).
\bibitem{CutlerFlanagan} C. Cutler and \'E.E. Flanagan,
        Phys. Rev. D {\bf 49}, 2658 (1994). 
\bibitem{MTW} See, for example, C.W. Misner, K.S. Thorne, 
        and J.A. Wheeler, {\it Gravitation} (Freeman, 
        San Francisco, 1973), Chapters 25 and 33.
\bibitem{BPT} J.M. Bardeen, W.H. Press, and S.A. Teukolsky,
        Astrophys. J. {\bf 178}, 347 (1972).
\bibitem{Shibataetal} M. Shibata, M. Sasaki, H. Tagoshi,
        and T. Tanaka, Phys. Rev. D {\bf 51}, 1646 (1995).
\bibitem{Ryan} F.D. Ryan, Phys. Rev. D {\bf 52}, R3159
        (1995).
\bibitem{Ori} A. Ori, Phys. Lett. {\bf A 202}, 347 (1995).
\bibitem{Will} For an overview, see 
        C.M. Will, in {\it Relativistic Cosmology},
        Proceedings of the Eighth Nishinomiya-Yukawa Memorial
        Symposium, edited by M. Sasaki (Universal Academy
        Press, Kyoto, Japan, 1994).
\bibitem{Shibata2} M. Shibata, Phys. Rev. D {\bf 48}, 663
        (1993); Prog. Theor. Phys. {\bf 90}, 595 (1993).
\bibitem{BransDicke} C. Brans and R.H. Dicke, Phys. Rev.
        {\bf 124}, 925 (1961).
\bibitem{WillBD} C.M. Will, Phys. Rev. D {\bf 50}, 6058 (1994).
\bibitem{Hawking} S.W. Hawking, Commun. Math. Phys. 
         {\bf 25}, 167 (1972).
\bibitem{Jackson} See, for example, J.D. Jackson, {\it Classical 
        Electrodynamics}, (Wiley, New York, 1975), p. 316. 
\bibitem{PoissonII} C. Cutler, L.S. Finn, E. Poisson,
        and G.J. Sussman, Phys. Rev. D {\bf 47}, 1511 (1993).
\bibitem{TagoshiNakamura} H. Tagoshi and T. Nakamura,
        Phys. Rev. D {\bf 49} 4016 (1994).
\bibitem{PoissonVI} E. Poisson, Phys. Rev. D {\bf 52}, 5719
        (1995).
\bibitem{Helstrom} C.W. Helstrom, {\it Statistical Theory of
        Signal Detection}, (Pergamon, Oxford, England, 1968).
\bibitem{Wainstein} L.A. Wainstein and V.D. Zubakov, 
        {\it Extraction of Signals from Noise}, (Prentice-Hall,
        Englewood Cliffs, 1962).
\bibitem{Thorne} K.S. Thorne, in {\it 300 Years of Gravitation}, 
        edited by S.W. Hawking and W. Israel (Cambridge University
        Press, Cambridge, England, 1987).
\bibitem{Schutz} B.F. Schutz, in {\it The detection of gravitational
        waves}, edited by D.G. Blair (Cambridge University
        Press, Cambridge, England, 1991).
\bibitem{PoissonWill} E. Poisson and C.M. Will, Phys. Rev.
        D {\bf 52}, 848 (1995).
\end{references}
\end{document}